\documentclass[final,5p,times,twocolumn]{elsarticle}

\usepackage{framed,multirow}

\usepackage{amssymb}
\usepackage{latexsym}

\usepackage{url}
\usepackage{xcolor}

\usepackage{hyperref}
\usepackage{amsmath} 
\usepackage{graphicx}
\usepackage[normalem]{ulem}
\useunder{\uline}{\ul}{}
\usepackage{comment}
\usepackage{booktabs}
\usepackage{float}
\usepackage{subcaption}

\definecolor{newcolor}{rgb}{.8,.349,.1}

\journal{Medical Image Analysis}

\begin{document}

\begin{frontmatter}

\title{NuLite - Lightweight and Fast Model for Nuclei Instance Segmentation and Classification}%

\author[1]{Cristian Tommasino\corref{cor1}}
\cortext[cor1]{Corresponding author}
\ead{cristian.tommasino@unina.it}
\author[1]{Cristiano Russo}
\ead{cristiano.russo@unina.it}
\author[1]{Antonio M. Rinaldi}
\ead{antoniomaria.rinaldi@unina.it}

\address[1]{Department of Electrical Engineering and Information Technology University of Naples Federico II, Via Claudio, 21, Naples 80125, Italy}

\begin{abstract}
%%%
In pathology, accurate and efficient analysis of Hematoxylin and Eosin (H\&E) slides is crucial for timely and effective cancer diagnosis. For these reasons, nuclei instance segmentation and classification tools are helpful, allowing pathologists to detect and identify regions of interest and perform quantitive analysis. Although many deep-learning solutions for this task exist in the literature, they often entail high computational costs and resource requirements, thus limiting their practical usage in medical applications.
To address this issue, we introduce NuLite, a U-Net-like architecture designed explicitly to be lightweight and fast. We obtained three versions of our model, NuLite-S, NuLite-M, and NuLite-H, trained on the PanNuke dataset. The experimental results prove that our models are equivalent to CellViT (SOTA) in terms of panoptic quality and F-score. However, our lightest model, NuLite-T, is about 58 times smaller in terms of parameters and about 10 times smaller in terms of GFlops. In comparison, our heaviest model is about 15 times smaller in terms of parameters and about 7 times smaller in terms of GFlops. Moreover, considering the GPU latency, our model is up to about 13 times faster than CellViT. Lastly, to prove the effectiveness of our solution, we provide a robust comparison of external datasets, namely CoNseP, MoNuSeg, and GlySAC. Our model is publicly available at \href{https://github.com/CosmoIknosLab/NuLite}{https://github.com/CosmoIknosLab/NuLite}.
%%%%
\end{abstract}


\end{frontmatter}

%\linenumbers

\section{Introduction}
\label{sec:1}
Cancer is a disease concerned with the uncontrolled growth and spread of abnormal cells, a significant global health challenge \cite{tran2022global}. Accurate diagnosis is essential in cancer treatment because it enables targeted therapies that improve patient outcomes and the chance of recovery. Advancements in computer vision techniques have significantly affected computational pathology (CPATH), opening new frontiers for analyzing histopathological images, like the Hematoxylin and Eosin (H\&E) stained one \cite{song2023artificial}. The precise segmentation and classification of cells became an exciting task in the literature due to the importance of understanding the morphology and topology of tissue in cancer diagnosis \cite{bahadir2024artificial}. However, this task in complex tissue environments poses many challenges due to the heterogeneity and overlap of nuclei structures, demanding robust and efficient solutions \cite{chen2024towards}.
To address these challenges, recent research has focused on developing sophisticated algorithms that leverage deep learning techniques, demonstrating superior performance in various image analysis tasks. These algorithms are designed to accurately identify and classify cellular components, even in complex and heterogeneous tissue environments. However, it is essential to note that these tools are meant to supplement pathologists and assist them in making more informed diagnostic decisions. Moreover, integrating machine learning models with domain-specific knowledge, such as the spatial relationships and morphological features of cells, has further enhanced the accuracy and robustness of computational pathology tools. This synergy between advanced computational methods and pathologists' expertise promises to significantly advance the field of cancer diagnostics, offering the potential for more personalized and effective treatment plans.

Over the years, scholars have proposed many methods to overcome traditional barriers encountered in histopathological analysis in different tasks \cite{van2021deep,tommasino2023histopathological,basu2024survey}. In particular, many deep learning solutions have shown promising results for nuclei instance segmentation and classification tasks, starting with the introduction of U-Net \cite{ronneberger2015u}. Furthermore, advanced neural network architectures, like ResNet \cite{he2016deep} and Vision Transformer (ViT) \cite{dosovitskiy2020image}, offered sophisticated mechanisms for learning detailed features and patterns without the constraints imposed by prior techniques, further improving the effectiveness of new models.
The recent trend toward integrating different modalities of deep learning, such as Convolutional Neural Networks (CNNs) combined with structures like U-Nets or multi-branch networks like HoVer-Net \cite{graham2019hover}, demonstrates the field's evolution toward more precise and robust techniques. Additionally, implementing spatial and morphological constraints within network architectures further refine their output, ensuring that cell segmentation is precise and contextually appropriate. Moreover, a recent technique, CellViT \cite{horst2024cellvit}, demonstrated using ViT to address nuclei instance segmentation and classification tasks, achieving the SOTA results.

This manuscript presents NuLite, a new UNet-like CNN \cite{ronneberger2015u} architecture designed for segmenting and classifying nuclei instances in Hematoxylin and Eosin (H\&E) images. Our architecture consists of the FastViT \cite{vasu2023fastvit} encoder, one decoder, and three segmentation heads purpose-built to perform one of the tasks identified in the HoVer-Net: nuclei prediction, horizontal and vertical map prediction, and nuclei classification \cite{graham2019hover}. We decided to use one decoder, contrary to what is commonly reported in the literature, to avoid parameter redundancy among the decoders and further reduce the parameters and GFLOPS.
NuLite is a faster, lighter alternative with state-of-the-art (SOTA) panoptic quality and detection performance. We proved its efficacy and efficiency through rigorous testing on benchmark datasets such as PanNuke \cite{gamper2020pannuke}. 
Furthermore, we conducted comprehensive evaluations on additional datasets such as MoNuSeg \cite{kumar2019multi}, CoNSeP \cite{graham2019hover}, and GlySAC \cite{doan2022sonnet}. NuLite consistently achieved SOTA results in these tests, outperforming advanced models like CellViT in various metrics, including precision, recall, and F1-score. These evaluations underscore the robustness and generalizability of NuLite across different types of histopathological images, highlighting its potential as a versatile tool in computational pathology.

Our main contributions to the field are significant regarding performance and its practical implications for enhancing diagnostic workflows. By enabling more accurate and efficient nuclei segmentation and classification, NuLite facilitates better quantitative analysis of tissue samples, which is crucial for improving diagnostic accuracy and patient outcomes in oncology and other medical disciplines.

We organized the rest of the paper as follows: Section \ref{sec:2} draws a brief state of the art about nuclei instance segmentation and vision transformer in pathology; Section \ref{sec:3} introduces our method, highlighting the architecture of the proposed CNN and loss function used to train it; Section \ref{sec:4} presents our experimental design and reports the experimental results with a comparison with SOTA and results on external datasets; lastly, Section \ref{sec:5} discusses the achieved results and Section \ref{sec:6} draws back the conclusions.

\section{Related Works}
\label{sec:2}
This section introduces the literature that addresses the nuclei instance segmentation and classification task. Then, we briefly introduce vision transform (ViT) literature.

\subsection{Nuclei Instance Segmentation}
Over the years, numerous methods for nuclei instance segmentation have been proposed. The first challenge they tried to overcome was to separate the overlapped nuclei; then, they addressed the classification of nuclei. In the following, we report the main work related to traditional and deep learning methods.

\subsubsection{Traditional methods}
In fluorescence microscopy, Malpica et al. \cite{malpica1997applying} proposed to use morphological watershed algorithms to effectively segment clustered nuclei, employing both gradient- and domain-based strategies to address the challenges of clustered nuclei segmentation. Similarly, Xiaodong Yang et al. \cite{yang2006nuclei} improved the tracking and analysis of nuclei in time-lapse microscopy via a marker-controlled watershed technique for initial segmentation, supplemented by mean-shift and Kalman filter techniques for dynamic and complex cellular behaviors.
Likewise, Jierong Cheng et al. \cite{cheng2008segmentation} improved segmentation accuracy by introducing shape markers derived from an adaptive H-minima transform associated with a marking function based on the outer distance transform. Stephan Wienert et al. \cite{wienert2012detection} involved a minimum-model strategy for the efficient detection and segmentation of cell nuclei in virtual microscopy images, simplifying the process while preserving effectiveness.
Instead, in histopathological imaging, the study by Afaf Tareef et al. \cite{tareef2018multi} introduced a multi-pass fast watershed method for accurate segmentation of overlapping cervical cells, using a novel three-pass process to segment both the nucleus and cytoplasm. Similarly, Miao Liao et al. \cite{liao2016automatic} developed a method that utilizes bottleneck detection and ellipse fitting to segment overlapping cells accurately.
Moreover, Sahirzeeshan Ali et al. \cite{ali2012integrated} provided a solution for overlapping objects in histological images by integrating region-based, boundary-based, and shape-based active contour models, significantly enhancing the segmentation accuracy of closely adjacent structures.
Instead, Veta et al. \cite{veta2013automatic} employed a marker-controlled watershed technique incorporating a multiscale approach and multiple marker types to improve nucleus segmentation in H\&E stained images for breast cancer histological images.

\subsubsection{Deep learning approaches}
In the last decade, deep learning techniques leveraged the limitations of traditional approaches. One of the first networks that achieved promising results in nuclei segmentation, posing the basis for all modern techniques, was U-Net proposed by Olaf Ronneberger et al. \cite{ronneberger2015u}. U-Net is an encoder-decoder neural network with skip connections, which helps preserve details crucial for medical image analysis. However, its original version proposed a way to separate clustered nuclei, which is a significant challenge in histopathology. 
Another network was BRP-Net \cite{song2016accurate} that creates nuclei proposals in the first place, then refines the boundary, and finally creates a segmentation out of this. However, this approach resulted in computationally intensive and slow.
Similarly, Alemi et al. introduced Mask-RCNN \cite{alemi2019nuclear}, built on Fast-RCNN \cite{he2017mask}, adding a segmentation branch after nuclei detection. Instead, Raza et al. proposed Micro-Net \cite{raza2019micro} updating U-Net to handle nuclei of varying sizes. Another network that significantly improved the nuclei instance segmentation and classification is HoVer-Net \cite{graham2019hover}, which has U-Net architecture with three branches that predict nuclei against the background, vertical and horizontal map, and nuclei types. The vertical and horizontal maps are crucial to separate overlapped nuclei and, in general, to perform instance segmentation. Following the idea of \cite{graham2019hover}, the authors in \cite{horst2024cellvit} proposed CellViT, which follows the same architecture but employs a ViT as the encoder, and the authors designed a decoder inspired by UNETR \cite{hatamizadeh2022unetr}. Instead, authors in \cite{tommasino2023hover} proposed a framework to obtain a smaller and lighter model than HoVerNet, HoVer-UNet, that is, a U-Net-like neural network with one decoder trained using a knowledge distillation approach.
Other recent networks proposed in the literature are STARDIST \cite{weigert2022nuclei}, and CPP-Net \cite{chen2023cpp}, which used star-convex polygons for segmentation, with CPP-Net enhancing the model by integrating shape-aware loss functions to improve accuracy.
Similarly, TSFD-Net \cite{ilyas2022tsfd} employed a Feature Pyramid Network and integrated a tissue-classifier branch to handle tissue-specific features, using advanced loss functions to manage class imbalance. Moreover, the SONNET \cite{doan2022sonnet} network is a deep learning model designed for simultaneous segmentation and classification of nuclei in large-scale multi-tissue histology images. It employs a self-guided ordinal regression approach that stratifies nuclear pixels based on their distance from the center of mass, improving the accuracy of segmenting overlapping nuclei. 

\subsection{Vision Transformers}
Vision Transformers (ViTs) have revolutionized image segmentation by providing advanced encoder-decoder architectures that enhance the capabilities of traditional U-Net-based models. Incorporating ViTs into these frameworks has enabled more precise instance and semantic segmentation across various domains, including medical imaging. TransUNet \cite{chen2021transunet} leverages a transformer to encode tokenized patches from CNN feature maps, effectively incorporating global context within the segmentation process. 
SETR \cite{zheng2021rethinking} uses the original ViT as the encoder and a fully convolutional network as the decoder, connected without intermediate skip connections, simplifying the architecture while maintaining performance.
UNETR \cite{hatamizadeh2022unetr} combining a standard ViT with a U-Net-like decoder that includes skipping connections, this model has shown to outperform others like TransUNet and SETR in medical image segmentation, demonstrating the effectiveness of integrating pre-trained ViTs with conventional segmentation networks.
Pre-training ViTs on large datasets is crucial for their success in segmentation tasks. Unlike CNNs, ViTs lack certain inductive biases and thus require substantial training data to learn effective representations. This is especially significant in medical imaging, where annotated data is limited. Self-supervised pre-training methods, such as DINO \cite{caron2021emerging}, have been pivotal in using unlabeled data to prime ViTs for fine-tuning specific segmentation tasks.
Xie et al. introduced Segformer \cite{xie2021segformer}, a model that utilizes a transformer as an image encoder coupled with a lightweight MLP decoder, focusing on efficiency and scalability.
FastViT \cite{vasu2023fastvit} is a high-speed hybrid vision transformer model that effectively balances latency and accuracy. It introduces a novel RepMixer component to reduce memory costs and enhance processing speed, making it faster and more efficient than traditional models across various image processing tasks.
\begin{figure*}[!h]
    \centering
    \includegraphics[width=\linewidth]{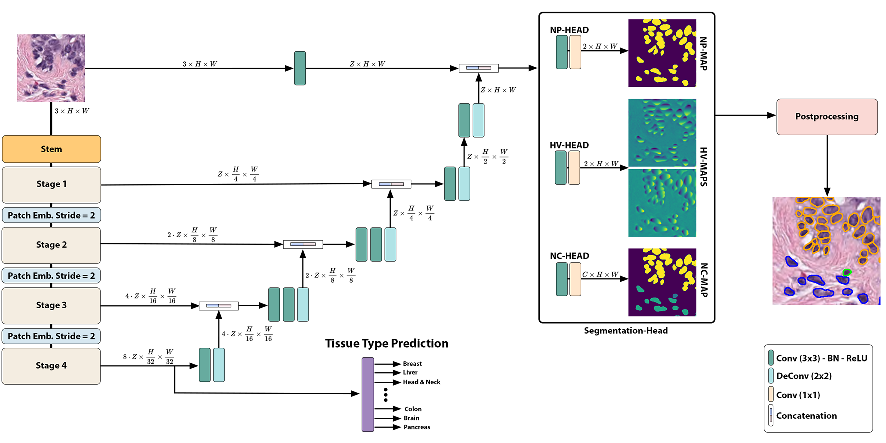}
    \caption{NuLite architecture. The network has a U-Net-like architecture with one decoder and three segmentation heads: one to predict nuclei, one to predict horizontal and vertical maps, and one to predict nuclei types. The post-processing uses all outputs to perform nuclei instance segmentation and assign the predicted nucleus type to each one.}
    \label{fig:arch}
\end{figure*}
\section{Methods}
\label{sec:3}
This section introduces NuLite architecture, the loss function used to train it, and the post-processing function. Lastly, we detail the inference pipeline.

\subsection{NuLite architecture} 
We designed NuLite with a U-Net-like architecture and three decoders, utilizing FastViT \cite{vasu2023fastvit}, a state-of-the-art Vision Transformer known for its lightweight and efficient design as the backbone. We decided to use U-Net architecture for its proven success in medical image segmentation tasks, as also shown by other recent models such as CellViT \cite{horst2024cellvit} and StartDist \cite{weigert2022nuclei}, which aligns with our goal of accurate nuclei detection and classification. Our approach draws inspiration from HoVer-Net \cite{graham2019hover}, which has established itself in the literature as an effective method for nuclei segmentation, as demonstrated by CellViT \cite{horst2024cellvit}, HoVer-NeXt \cite{baumann2024hover}, and HoVer-UNet \cite{tommasino2023hover}, but using just one decoder with three segmentation heads. Therefore, our model predicts nuclei maps, type maps, and horizontal and vertical maps, followed by a watershed algorithm to perform nuclei instance segmentation in a postprocessing step. Thus, our network comprises three heads: the NP-HEAD for nuclei segmentation against the background, the HV-HEAD for predicting horizontal and vertical orientation maps, and the NC-HEAD for nuclei classification, as illustrated in Figure \ref{fig:arch}. This architecture supports detailed nuclei analysis through a postprocessing step that leverages the NP-MAP and HV-MAPS to precisely detect individual nuclei, subsequently using the NC-MAP to assign the type to each nucleus instance. We carefully designed the decoders to minimize computational overhead, maintaining low parameter counts and GFlops, thus ensuring efficiency. Furthermore, we integrated a dense layer within the encoder to facilitate tissue classification, extending the functionality of the network beyond nuclei analysis, and results useful during the training step to improve the segmentation and classification capabilities.

We focused on the decoder design and built it to work with the FastViT \cite{vasu2023fastvit} encoder. Therefore, the decoder consists of five main layers and three segmentation heads, as detailed in Table \ref{tab:arch_det}. 
\begin{table}[!h]
    \centering
    \caption{NuLite decoder details}
    \label{tab:arch_det}
\resizebox{\linewidth}{!}{%
    \begin{tabular}{lp{4.5cm}ll}
            \toprule
         \#Layer & Layer composition & Input Shape & Output shape \\
         \midrule
          DEC.1& Conv2D (3x3) - BN - ReLU \newline DeConv (2x2)  & $8\cdot Z \times \frac{H}{32} \times \frac{W}{32}$ & $4\cdot Z \times \frac{H}{16} \times \frac{W}{16}$\\
          DEC.2& Conv2D (3x3) - BN - ReLU \newline Conv2D (3x3) - BN - ReLU\newline DeConv (2x2)& $8\cdot Z \times \frac{H}{16} \times \frac{W}{16}$& $2\cdot Z \times \frac{H}{8} \times \frac{W}{8}$\\
          DEC.3&  Conv2D (3x3) - BN - ReLU \newline Conv2D (3x3) - BN - ReLU\newline DeConv (2x2)& $4\cdot Z \times \frac{H}{8} \times \frac{W}{8}$& $Z \times \frac{H}{4} \times \frac{W}{4}$\\
          DEC.4&  Conv2D (3x3) - BN - ReLU \newline DeConv (2x2) & $2 \cdot Z \times \frac{H}{4} \times \frac{W}{4}$& $Z \times \frac{H}{2} \times \frac{W}{2}$\\
          DEC.5&  Conv2D (3x3) - BN - ReLU \newline DeConv (2x2) & $Z \times \frac{H}{2} \times \frac{W}{2}$& $Z \times H \times W$\\
          NP.HEAD&  Conv2D (3x3) - BN - ReLU \newline Conv2D (1x1) & $2 \cdot Z \times H \times W$& $2 \times H \times W$\\
          HV.HEAD &  Conv2D (3x3) - BN - ReLU \newline Conv2D (1x1) & $2 \cdot Z \times H \times W$& $2 \times H \times W$\\
          NC.HEAD &  Conv2D (3x3) - BN - ReLU \newline Conv2D (1x1) & $2 \cdot Z \times H \times W$& $C \times H \times W$\\
         \bottomrule
    \end{tabular}}
\end{table}
As a standard U-like architecture, we employ the skip connection between the main block output of the encoder, namely stage 1 to stage 4, as shown in Figure \ref{fig:arch}, and each main block of our decoder. Furthermore, we add a skip connection between the original input after a convolutional layer and the last layer of the decoder. 
The decoder architecture comprises five layers, namely DEC.1, DEC.2, DEC.3, DEC.4, and DEC.5, and three segmentation heads, namely NP.HEAD, HV.HEAD, and NC.HEAD, as described in Table \ref{tab:arch_det}. These layers operate on input images defined by dimensions height \( H \) and width \( W \), with \( Z \) indicating the number of channels output from the encoder.
We structured the DEC.1, DEC.4, and DEC.5 with a \(3 \times 3\) convolutional layer, which is succeeded by batch normalization and ReLU activation and augmented by a deconvolution layer. The output from DEC.1 and DEC.4 yields feature maps with half the number of channels of \( Z \) but with dimensions expanded to twice the height (\(2H\)) and width (\(2W\)). However, DEC.5 maintains the channel count of \( Z \) and doubles the height and width.
The design of DEC.2 and DEC.3 integrates two \(3 \times 3\) convolutional layers, each followed by batch normalization and ReLU activation. These layers are completed by a deconvolution layer that produces outputs with a quarter of the channels of \( Z \) and twice the original dimensions in height and width.
Each segmentation head has a \(3 \times 3\) convolution followed by batch normalization and ReLU, followed by a \(1 \times 1\) convolution that adjusts the output channels to meet specific requirements, namely 2 channels for nuclei prediction and horizontal and vertical map prediction and $C$ channels for nuclei classification where $C$ is the number of class contained in the training dataset. 
As notable from the decoder structure, the second and third layers contain two convolutional blocks, while the rest have only one covolutional block. That is because the number of feature maps is reduced by 4 times in the second and third and 2 times in the rest.
FastViT exists in several configurations, including T8, T12, S12, SA12, SA24, SA36, and MA36. The parameter \( Z \), which denotes the number of channels, varies across these models. Specifically, \( Z \) is set to 384 for T8, while it remains consistent at 512 for T12, S12, SA12, SA24, and SA36. For the MA36 configuration, \( Z \) increases to 608. Therefore, in this paper, we consider a server version of NuLite, each using a version of FastViT.

\subsection{Loss fuction}
To train NuLite, we use a combination of different loss functions for each network output, as also suggested in \cite{graham2019hover, horst2024cellvit}. Therefore, the total loss is defined as the sum of a loss for each segmentation head, as shown in Equation \ref{eq:total_loss}.
\begin{equation}
    \label{eq:total_loss}
    L_{\text{total}} = L_{NP} + L_{HV} + L_{NT} + L_{TC}
\end{equation}

$L_{NP}$ is the loss for the NP-HEAD, defined as a linear combination of Focal Tversky loss (FTL) and Dice loss (DICE), as shown in Equation \ref{eq:loss_np}.
\begin{equation}
    \label{eq:loss_np}
    L_{NP} = \lambda_{NP}^{\text{FTL}} L_{\text{FTL}} + \lambda_{NP}^{\text{DICE}} L_{\text{DICE}}
\end{equation}
$L_{HV}$ is the loss for the HV-HEAD, defined as a linear combination of Mean Square Error (MSE) and Mean Square Gradient Error (MSGE), as shown in Equation \ref{eq:loss_hv}. 
\begin{equation}
    \label{eq:loss_hv}
    L_{HV} = \lambda_{HV}^{\text{MSE}} L_{\text{MSE}} + \lambda_{HV}^{\text{MSGE}} L_{\text{MSGE}}
\end{equation}
$L_{nt}$ is the loss for the NT-HEAD, defined as a linear combination of FLT, DICE, and Binary Cross Entropy Loss (BCE) as shown in Equation \ref{eq:loss_nt}.
\begin{equation}
    \label{eq:loss_nt}
    L_{NT} = \lambda_{NT}^{\text{FTL}} L_{\text{FTL}} + \lambda_{NT}^{\text{DICE}} L_{\text{DICE}} + \lambda_{NT}^{\text{BCE}} L_{\text{BCE}}
\end{equation}
$L_{TC}$ the loss for the tissue classification, computed as Cross Entropy (CE) as shown in Equation \ref{eq:loss_tc}
\begin{equation}
    \label{eq:loss_tc}
    L_{TC} = \lambda_{TC}^{\text{CE}} L_{\text{CE}}
\end{equation}
In these equations, $\lambda_{\text{brach}^{\text{loss}}}$ coefficients represent the weight given to each loss component.
\subsection{Post-Processing}
As described in the preview sections, our network, NuLite, encloses three specialized segmentation heads dedicated to extracting essential information for a nuclei instance segmentation and classification postprocessing step. 
Due to our network following the idea proposed in HoVer-Net and Cell-ViT, postprocessing is a crucial step in refining the raw predictions produced by the network. NP-HEAD output is a probability map indicating the likelihood of each pixel belonging to a nucleus. A threshold is applied to this probability map to generate a binary mask. Pixels with probabilities above the threshold are considered part of a nucleus. HV-HEAD contains horizontal and vertical gradient maps (HV maps) that help to delineate the nuclei boundaries more accurately. These maps provide additional information about the direction and magnitude of changes in the image, which is useful for refining the edges of the segmented nuclei. The gradient information from the HV maps is used to split merged nuclei. This process is critical when there are overlapped nuclei. Each nucleus identified from the segmentation step is classified according to the output of NC-HEAD.  
Finally, some morphological operations, like dilation and erosion, can be applied to smooth the boundaries of the segmented nuclei and remove small noise artifacts, improving the visual quality of the segmentation masks.
\begin{figure*}[ht!]
    \centering
    \includegraphics[width=\linewidth]{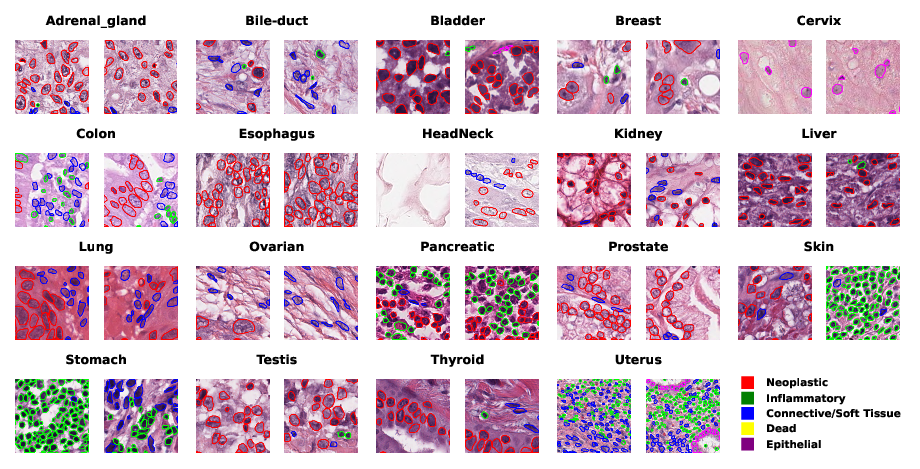}
    \caption{Example images from the PanNuke dataset showing varied tissue types and nuclear annotations.}
    \label{fig:pannuke}
\end{figure*}
\section{Experimental Results}
\label{sec:4}
This section introduces the dataset employed, the metrics used to evaluate our network, the training details, and the experimental results with a related comparison with SOTA on the PanNuke dataset. Lastly, we comprehensively analyze inference time and network complexity and show the results on another external dataset.
\subsection{Datasets}

\paragraph{PanNuke}
The PanNuke dataset \cite{gamper2020pannuke} is the primary resource for training and evaluating our model. It comprises 189,744 annotated nuclei across 7,904 images, each of size 256$\times$256 pixels, spanning 19 distinct tissue types and categorized into five unique cell classes. These cell images were captured at a 40$\times$ magnification with a fine resolution of 0.25 $\mu$m/px. Notably, the dataset exhibits a significant class imbalance; particularly, the nuclei class of dead cells is markedly underrepresented, evident from the nuclei and tissue class statistics.

\begin{figure*}[!ht]
     \centering
     \begin{subfigure}[b]{1\linewidth}
         \centering
         \includegraphics[width=\linewidth]{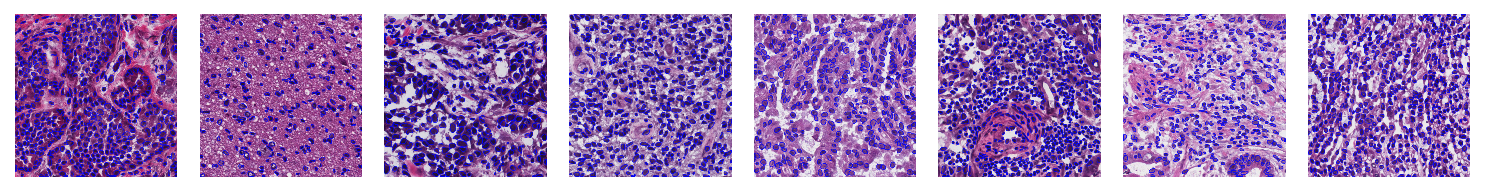}
         \caption{MoNuSeg}
         \label{fig:monuseg}
     \end{subfigure}
     \hfill
     \begin{subfigure}[b]{1\linewidth}
         \centering
         \includegraphics[width=\linewidth]{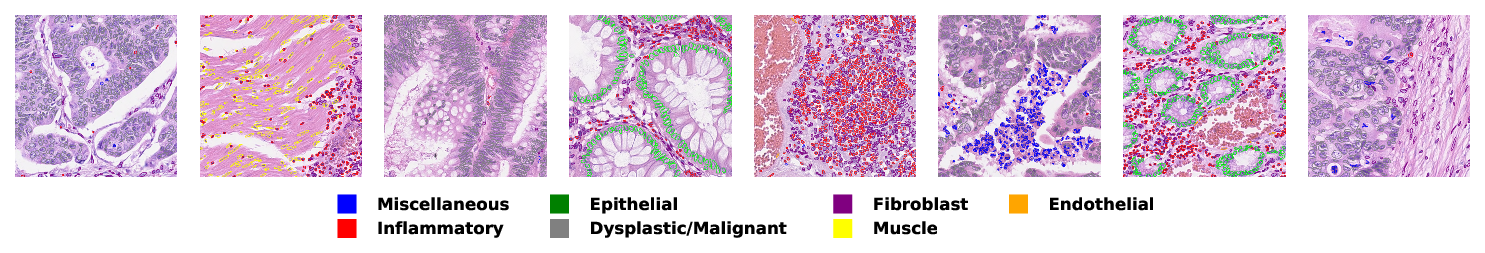}
         \caption{CoNSeP}
         \label{fig:tconsep}
     \end{subfigure}
     \hfill
     \begin{subfigure}[b]{1\linewidth}
         \centering
         \includegraphics[width=\linewidth]{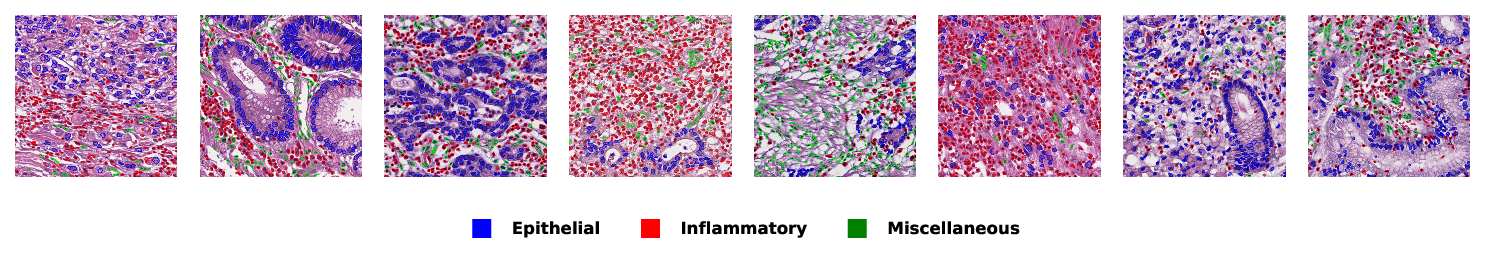}
         \caption{GlySAC}
         \label{fig:glysac}
     \end{subfigure}
        \caption{Examples images from MoNuSeg, CoNSeP, and GlySAC dataset with annotations}
        \label{fig:dataset_examples}
\end{figure*}
\paragraph{MoNuSeg}
The MoNuSeg dataset \cite{kumar2019multi} is employed as a supplementary resource for nuclei segmentation. Unlike PanNuke, MoNuSeg is considerably smaller and does not categorize nuclei into various classes. In this study, only the test subset of MoNuSeg is used to assess our model. This subset includes 14 high-resolution images (1000 $\times$ 1000 px) captured at 40$\times$ magnification and a resolution of 0.25 $\mu$m/px, containing over 7,000 annotated nuclei spanning seven organ types (kidney, lung, colon, breast, bladder, prostate, and brain) across various disease states. Due to the absence of nuclei labels, classification performance cannot be evaluated with this dataset.

\paragraph{CoNSeP}
The CoNSeP dataset \cite{graham2019hover}, curated by Graham et al., comprises 41 H\&E-stained colorectal adenocarcinoma whole slide images (WSIs) at a resolution of 0.25 $\mu$m/px, resized to 1024 $\times$ 1024 px to facilitate processing. This diverse dataset features stromal, epithelial, muscular, collagen, adipose, and tumorous regions. It also includes a variety of nuclei types derived from different originating cells, such as normal epithelial, dysplastic epithelial, inflammatory, necrotic, muscular, fibroblast, and miscellaneous nuclei types, including necrotic and mitotic cells, which aids in comprehensive phenotypic analysis.

\paragraph{GlySAC}
The GLySAC dataset \cite{doan2022sonnet}, short for Gastric Lymphocyte Segmentation and Classification, focuses on segmenting and classifying nuclei within gastric pathology. It contains 59 H\&E stained image tiles, each 1000x1000 pixels, sourced from gastric adenocarcinoma WSIs and captured at a 40$\times$ magnification using an Aperio digital scanner. The dataset encapsulates a total of 30,875 nuclei, categorized into three primary groups: Lymphocytes (12,081 nuclei), Epithelial nuclei (12,287 nuclei), encompassing both cancerous and normal cells, and Miscellaneous other nuclei types (6,507 nuclei).

\subsection{Evaluation metrics}
In evaluating nuclear instance segmentation, traditional metrics such as the Dice coefficient and Jaccard index often fall short as they do not adequately reflect the detection quality of individual nuclei or the precision in segmenting overlapping nuclei. Therefore, more sophisticated metrics are employed as suggested in \cite{graham2019hover, horst2024cellvit, doan2022sonnet, ilyas2022tsfd}.
\paragraph{Panoptic Quality}
The Panoptic Quality (PQ) metric provides a comprehensive evaluation by combining two essential aspects given by Detection Quality (DQ) and Segmentation Quality (SQ). DQ reflects how well the model detects and correctly identifies individual nuclei, calculated as denoted in Equation \ref{eq:dq}, where $TP$, $FP$, and $FN$ represent the true positives, false positives, and false negatives, respectively.
\begin{equation}
    \label{eq:dq}
    DQ = \frac{|TP|}{|TP| + \frac{1}{2}|FP| + \frac{1}{2}|FN|}
\end{equation}
SQ assesses the accuracy of the segmentation for the detected nuclei, computed as the mean IoU (Intersection over Union) of matched pixels, as denoted in Equation \ref{eq:sq}, where $y$ and $\hat{y}$ denote the ground truth and predicted segments, respectively.
\begin{equation}
    \label{eq:sq}
    SQ = \frac{\sum_{(y, \hat{y}) \in TP} IoU(y, \hat{y})}{|TP|}
\end{equation}
Therefore, PQ is the product of detection and segmentation quality, as denoted in equation \ref{eq:pq}.
\begin{equation}
    \label{eq:pq}
    PQ = DQ \times SQ
\end{equation}

In this work, we consider two adaptions of PQ: Binary PQ (bPQ), which considers all nuclei as a single class against the background, and Multi-class PQ (mPQ), which Evaluates PQ separately for each class of nuclei and averages the scores.

\paragraph{F1-score}
Several metrics commonly utilized in machine learning were employed to evaluate instance classification performance. Precision (\(P\)), which quantifies the accuracy of the positive predictions, is defined in the Equation \ref{eq:prec}.
\begin{equation}
P = \frac{TP}{TP + FP}
\label{eq:prec}
\end{equation}
Where \(TP\) represents true positives and \(FP\) represents false positives. Recall (\(R\)), also known as sensitivity, measures the ability of the model to detect all relevant instances, defined in Equation \ref{eq:recall}
\begin{equation}
R = \frac{TP}{TP + FN}
\label{eq:recall}
\end{equation}
With \(FN\) indicating false negatives. The F1 Score, a harmonic mean of precision and recall that balances these metrics is crucial in uneven class distribution, shown in Equation \ref{eq:f1}.
\begin{equation}
F1 = 2 \times \frac{P \times R}{P + R}
\label{eq:f1}
\end{equation}
Accuracy, indicating the overall correctness of the model, is formulated in Equation \ref{eq:acc}:
\begin{equation}
Accuracy = \frac{TP + TN}{TP + TN + FP + FN}
\label{eq:acc}
\end{equation}
where \(TN\) represents true negatives.

To detail performance assessment in multi-class settings, the F1 Score is further refined through equations that include terms for each class \(c\) and \(d\), illustrating both traditional components and inter-class effects, shown in Equations \ref{eq:p_c}, \ref{eq:r_c}, and \ref{eq:f_c}

\begin{equation}
P_c = \frac{T_{Pc} + T_{Nc}}{T_{Pc} + T_{Nc} + 2FP_c + FP_d}
\label{eq:p_c}
\end{equation}
\begin{equation}
R_c = \frac{T_{Pc} + T_{Nc}}{T_{Pc} + T_{Nc} + 2FN_c + FN_d}
\label{eq:r_c}
\end{equation}
\begin{equation}
F1_c = \frac{2(T_{Pc} + T_{Nc})}{2(T_{Pc} + T_{Nc}) + 2FP_c + 2FN_c + FP_d + FN_d}
\label{eq:f_c}
\end{equation}

\subsection{Results on PanNuke}
In this subsection, we detail the training strategy used to train NuLite on PanNuke and show its experimental results compared to similar methods.

\paragraph{Training}
We used the AdamW optimizer for the training set, configured with specific hyperparameters, including beta values of 0.85 and 0.95, a learning rate of 0.0003, and a weight decay of 0.0001. An exponential scheduler managed the learning rate decay with a gamma of 0.85, effectively adjusting the learning rate across the epochs. Furthermore, we trained the model with a batch size 16 for 130 epochs. 
We used data augmentation techniques to ensure the model generalized well across different imaging conditions. We employed geometric transformations, including rotations, flips, elastic transformations, simulated cell orientations, and position variations; photometric transformations, including blur, Gaussian noise, color jitter, and superpixel augmentation; and enhanced robustness against variations in stain quality and imaging noise. 
Lastly, we used a specific sampling strategy focusing on cell and tissue types, ensuring a balanced representation of various classes in the training batches, as shown in \cite{horst2024cellvit}.

\paragraph{Training Results}
\begin{table}[!h]
\centering
\caption{Average PQ across the three PanNuke splits for each
nucleus type on the PanNuke dataset. The best results are highlighted in bold, with the second-best in underlined text.}
\label{tab:pq_pannuke}
\resizebox{\columnwidth}{!}{%
\begin{tabular}{llllll}
\toprule
Model                        & Neoplastic & Epithelial & Inflammatory & Connective & Dead  \\
\midrule
DIST          & 0.4390          & 0.2900          & 0.3430          & 0.2750          & 0.0000          \\
Mask-RCNN     & 0.4720          & 0.4030          & 0.2900          & 0.3000          & 0.0690          \\
Micro-Net     & 0.5040          & 0.4420          & 0.3330          & 0.3340          & 0.0510          \\
HoVer-Net     & 0.5510          & 0.4910          & 0.4170          & 0.3880          & 0.1390          \\
HoVer-UNet    & 0.5240          & 0.4780          & 0.4010          & 0.3790          & 0.0760          \\
CellViT256    & 0.5670          & 0.5590          & 0.4050          & 0.4050          & 0.1440          \\
CellViT-SAM-H & \textbf{0.5810} & \textbf{0.5830} & 0.4170          & \textbf{0.4230} & \textbf{0.1490} \\
\midrule
NuLite-T      & 0.5722          & 0.5622          & 0.4155          & 0.4062          & 0.1370          \\
NuLite-M      & 0.5752          & 0.5693          & \textbf{0.4308} & 0.4070          & 0.1379          \\
NuLite-H      & {\ul 0.5765}    & {\ul 0.5712}    & {\ul 0.4171}    & {\ul 0.4134}    & {\ul 0.1447}    \\
\bottomrule
\end{tabular}%
}
\end{table}
\begin{table*}[!h]
	\centering
	\caption{Precision (P), Recall (R), and F1-score (F1) across the three PanNuke splits for binary detection and each nucleus type. The best results are highlighted in bold, with the second-best in underlined text.}
	\label{tab:f1_pannuke}
	\resizebox{\linewidth}{!}{%
            \begin{tabular}{llllllllllllllllllllllllll}
\toprule
			\multicolumn{1}{l}{Model} && \multicolumn{3}{l}{Detection} & \multicolumn{1}{c}{}  & \multicolumn{19}{l}{Classification}\\    
			\cmidrule{7-25}
			&& \multicolumn{3}{c}{} & \multicolumn{1}{c}{} & \multicolumn{3}{l}{Neoplastic} & \multicolumn{1}{c}{} & \multicolumn{3}{l}{Epithelial} & \multicolumn{1}{c}{} & \multicolumn{3}{l}{Inflammatory} & \multicolumn{1}{l}{} & \multicolumn{3}{l}{Connective} & \multicolumn{1}{l}{} & \multicolumn{3}{l}{Dead} \\
			\cmidrule{7-9} \cmidrule{11-13} \cmidrule{15-17} \cmidrule{19-21} \cmidrule{23-25}
			           &        & $P$              & $R$              & $F1$             &   & $P$              & $R$              & $F1$             &   & $P$              & $R$              & $F1$             &   & $P$              & $R$              & $F1$             &   & $P$              & $R$           & $F1$             &           & $P$              & $R$              & $F1$             \\
			\midrule
DIST          &  & 0.74          & 0.71          & 0.73          & {\ul } & 0.49          & 0.55          & 0.50          & {\ul }               & 0.38          & 0.33          & 0.35          & {\ul }               & 0.42          & 0.45          & 0.42          & {\ul }               & 0.42          & 0.37          & 0.39          & {\ul }               & 0.00          & 0.00          & 0.00          \\
Mask-RCNN     &  & 0.76          & 0.68          & 0.72          & {\ul } & 0.55          & 0.63          & 0.59          & {\ul }               & 0.52          & 0.52          & 0.52          & {\ul }               & 0.46          & 0.54          & 0.50          & {\ul }               & 0.42          & 0.43          & 0.42          & {\ul }               & 0.17          & 0.30          & 0.22          \\
Micro-Net     &  & 0.78          & \textbf{0.82} & 0.80          & {\ul } & 0.59          & 0.66          & 0.62          & {\ul }               & 0.63          & 0.54          & 0.58          & {\ul }               & {\ul 0.59}    & 0.46          & 0.52          & {\ul }               & 0.50          & 0.45          & 0.47          & {\ul }               & 0.23          & 0.17          & 0.19          \\
HoVer-Net     &  & 0.82          & 0.79          & 0.80          & {\ul } & 0.58          & 0.67          & 0.62          & {\ul }               & 0.54          & 0.60          & 0.56          & {\ul }               & 0.56          & 0.51          & 0.54          & {\ul }               & 0.52          & 0.47          & 0.49          & {\ul }               & 0.28          & \textbf{0.35} & 0.31          \\
HoVer-UNet    &  & 0.80          & 0.79          & 0.79          & {\ul } & 0.59          & 0.69          & 0.64          & {\ul }               & 0.57          & 0.67          & 0.62          & {\ul }               & 0.55          & 0.52          & 0.53          & {\ul }               & 0.52          & 0.45          & 0.48          & {\ul }               & 0.21          & 0.16          & 0.18          \\
CellViT256    &  & {\ul 0.83}    & \textbf{0.82} & {\ul 0.82}    & {\ul } & 0.69          & {\ul 0.70}    & 0.69          & {\ul }               & 0.68          & 0.71          & 0.70          & {\ul }               & {\ul 0.59}    & \textbf{0.58} & \textbf{0.58} & {\ul }               & 0.53          & {\ul 0.51}    & {\ul 0.52}    & {\ul }               & 0.39          & \textbf{0.35} & \textbf{0.37} \\
CellViT-SAM-H &  & \textbf{0.84} & {\ul 0.81}    & \textbf{0.83} & {\ul } & \textbf{0.72} & 0.69          & \textbf{0.71} & {\ul }               & \textbf{0.72} & 0.73          & \textbf{0.73} & {\ul }               & {\ul 0.59}    & {\ul 0.57}    & \textbf{0.58} & {\ul }               & \textbf{0.55} & \textbf{0.52} & \textbf{0.53} & {\ul }               & 0.43          & {\ul 0.32}    & {\ul 0.36}    \\
\midrule
NuLite-T      &  & 0.82          & \textbf{0.82} & {\ul 0.82}    & {\ul } & 0.68          & \textbf{0.71} & {\ul 0.70}    & {\ul }               & \textbf{0.72} & 0.72          & {\ul 0.72}    & {\ul }               & {\ul 0.59}    & 0.56          & {\ul 0.57}    & {\ul }               & 0.52          & \textbf{0.52} & {\ul 0.52}    & {\ul }               & {\ul 0.44}    & 0.30          & {\ul 0.36}    \\
NuLite-M      &  & {\ul 0.83}    & \textbf{0.82} & \textbf{0.83} & {\ul } & {\ul 0.70}    & \textbf{0.71} & {\ul 0.70}    & {\ul }               & {\ul 0.71}    & \textbf{0.75} & \textbf{0.73} & {\ul }               & 0.58          & \textbf{0.58} & \textbf{0.58} & {\ul }               & {\ul 0.54}    & {\ul 0.51}    & {\ul 0.52}    & {\ul }               & \textbf{0.48} & 0.30          & \textbf{0.37} \\
NuLite-H      &  & {\ul 0.83}    & \textbf{0.82} & \textbf{0.83} & {\ul } & {\ul 0.70}    & \textbf{0.71} & \textbf{0.71} & {\ul }               & \textbf{0.72} & {\ul 0.74}    & \textbf{0.73} & {\ul }               & \textbf{0.60} & {\ul 0.57}    & \textbf{0.58} & {\ul }               & {\ul 0.54}    & \textbf{0.52} & \textbf{0.53} & {\ul }               & \textbf{0.48} & 0.30          & \textbf{0.37}   \\
\bottomrule
\end{tabular}%
	}
\end{table*}
\begin{table*}[h!]
\centering
\caption{Multi-class Panoptic Quality (mPQ) and binary Panoptic Quality (bPQ) across the three PanNuke splits over tissues among HoVerNet, CellViT, and NuLite. The best results are highlighted in bold, with the second-best in underlined text.}
\label{tab:tissue_res}
\resizebox{\linewidth}{!}{%
\begin{tabular}{lllllllllllllllllll}
\toprule
Tissue       &  & \multicolumn{2}{l}{HoVer-Net} & \multicolumn{1}{l}{} & \multicolumn{2}{l}{CellViT256} & \multicolumn{1}{l}{} & \multicolumn{2}{l}{CellViT-SAM-H} & \multicolumn{1}{c}{} & \multicolumn{2}{l}{NuLite-T} & \multicolumn{1}{c}{} & \multicolumn{2}{l}{NuLite-M} & \multicolumn{1}{c}{} & \multicolumn{2}{l}{NuLite-H} \\
\cmidrule{3-4} \cmidrule{6-7} \cmidrule{9-10} \cmidrule{12-13} \cmidrule{15-16} \cmidrule{18-19}
             &  & mPQ                & bPQ      &                      & mPQ            & bPQ           &                      & mPQ             & bPQ             &                      & mPQ               & bPQ           &                      & mPQ               & bPQ            &                      & mPQ              & bPQ             \\
\midrule
Adrenal      &  & 0.481              & 0.696    &                      & 0.495          & 0.701         &                      & \textbf{0.513} & {\ul 0.709}          &                      & 0.503             & 0.707         &                      & 0.500          & \textbf{0.712}       &                      & {\ul 0.511}    & 0.706                \\
Bile Duct    &  & 0.471              & 0.670    &                      & 0.472          & 0.671         &                      & \textbf{0.489} & \textbf{0.678}       &                      & 0.477             & 0.671         &                      & {\ul 0.483}    & {\ul 0.674}          &                      & 0.480          & 0.672                \\
Bladder      &  & 0.579              & 0.703    &                      & 0.576          & 0.706         &                      & {\ul 0.584}    & 0.707                &                      & 0.571             & 0.704         &                      & 0.582          & \textbf{0.720}       &                      & \textbf{0.586} & {\ul 0.719}          \\
Breast       &  & 0.490              & 0.647    &                      & 0.509          & {\ul 0.664}   &                      & \textbf{0.518} & \textbf{0.675}       &                      & 0.507             & 0.660         &                      & 0.507          & {\ul 0.664}          &                      & {\ul 0.510}    & 0.663                \\
Cervix       &  & 0.444              & 0.665    &                      & 0.489          & 0.686         &                      & 0.498          & 0.687                &                      & 0.493             & 0.683         &                      & \textbf{0.508} & {\ul \textbf{0.693}} &                      & {\ul 0.502}    & {\ul \textbf{0.693}} \\
Colon        &  & 0.410              & 0.558    &                      & 0.425          & 0.570         &                      & \textbf{0.449} & \textbf{0.592}       &                      & 0.434             & 0.573         &                      & {\ul 0.445}    & 0.582                &                      & 0.443          & {\ul 0.584}          \\
Esophagus    &  & 0.509              & 0.643    &                      & 0.537          & 0.662         &                      & {\ul 0.545}    & 0.668                &                      & 0.528             & 0.661         &                      & 0.543          & {\ul 0.673}          &                      & \textbf{0.554} & \textbf{0.675}       \\
Head \& Neck &  & 0.453              & 0.633    &                      & 0.490          & 0.647         &                      & 0.491          & \textbf{0.654}       &                      & \textbf{0.494}    & 0.645         &                      & {\ul 0.492}    & {\ul 0.652}          &                      & 0.491          & 0.645                \\
Kidney       &  & 0.442              & 0.684    &                      & {\ul 0.541}    & 0.699         &                      & 0.537          & \textbf{0.709}       &                      & 0.533             & 0.698         &                      & 0.540          & {\ul 0.705}          &                      & \textbf{0.545} & 0.701                \\
Liver        &  & 0.497              & 0.725    &                      & 0.507          & 0.716         &                      & {\ul 0.522}    & 0.732                &                      & 0.512             & 0.724         &                      & 0.517          & \textbf{0.734}       &                      & \textbf{0.523} & {\ul 0.733}          \\
Lung         &  & 0.400              & 0.630    &                      & 0.410          & 0.632         &                      & {\ul 0.431}    & {\ul \textbf{0.643}} &                      & 0.417             & 0.630         &                      & 0.419          & {\ul \textbf{0.643}} &                      & \textbf{0.432} & {\ul \textbf{0.643}} \\
Ovarian      &  & 0.486              & 0.631    &                      & 0.526          & 0.660         &                      & {\ul 0.539}    & \textbf{0.672}       &                      & 0.529             & 0.665         &                      & \textbf{0.540} & 0.667                &                      & 0.537          & {\ul 0.671}          \\
Pancreatic   &  & 0.460              & 0.649    &                      & 0.477          & 0.664         &                      & 0.472          & 0.666                &                      & 0.485             & 0.665         &                      & \textbf{0.487} & {\ul \textbf{0.677}} &                      & {\ul 0.486}    & {\ul \textbf{0.677}} \\
Prostate     &  & 0.510              & 0.662    &                      & 0.516          & 0.670         &                      & \textbf{0.532} & \textbf{0.682}       &                      & 0.514             & 0.667         &                      & 0.519          & {\ul 0.676}          &                      & {\ul 0.529}    & 0.674                \\
Skin         &  & 0.343              & 0.623    &                      & 0.366          & 0.640         &                      & \textbf{0.434} & \textbf{0.657}       &                      & {\ul 0.422}       & 0.642         &                      & 0.421          & {\ul 0.649}          &                      & 0.406          & 0.636                \\
Stomach      &  & \textbf{0.473}     & 0.689    &                      & 0.448          & 0.692         &                      & {\ul 0.471}    & {\ul 0.702}          &                      & 0.455             & 0.695         &                      & 0.465          & \textbf{0.706}       &                      & 0.454          & 0.698                \\
Testis       &  & 0.475              & 0.689    &                      & 0.509          & 0.688         &                      & 0.513          & {\ul 0.696}          &                      & 0.500             & 0.682         &                      & \textbf{0.528} & 0.691                &                      & {\ul 0.517}    & \textbf{0.697}       \\
Thyroid      &  & 0.432              & 0.698    &                      & 0.441          & 0.704         &                      & 0.452          & \textbf{0.715}       &                      & \textbf{0.459}    & 0.708         &                      & 0.448          & 0.707                &                      & {\ul 0.454}    & {\ul 0.710}          \\
Uterus       &  & 0.439              & 0.639    &                      & {\ul 0.474}    & 0.652         &                      & {\ul 0.474}    & \textbf{0.663}       &                      & 0.469             & 0.648         &                      & \textbf{0.488} & {\ul 0.660}          &                      & 0.473          & 0.658                \\
\midrule
Average      &  & 0.463              & 0.660    &                      & 0.485          & 0.670         &                      & \textbf{0.498} & \textbf{0.679}       &                      & 0.490             & 0.670         &                      & {\ul 0.496}    & {\ul 0.678}          &                      & {\ul 0.496}    & 0.677                \\
STD          &  & 0.050              & 0.038    &                      & 0.050          & 0.034         &                      & {\ul 0.041}    & \textbf{0.032}       &                      & \textbf{0.040}    & {\ul 0.035}   &                      & 0.043          & {\ul 0.035}          &                      & 0.046          & {\ul 0.035}         
\\
\bottomrule
\end{tabular}%
}
\end{table*}
\begin{table*}[!ht]
\centering
\caption{Comparison between NuLite using reparameterized or no-reparameterized FastViT as an encoder in terms of binary panoptic quality (bPQ), multiclass panoptic quality (mPQ), PQ for each nucleus type, F1-score, and F1-score for each nucleus type.}
\label{tab:ablation}
\resizebox{\linewidth}{!}{%
\begin{tabular}{clllllllllllllllll}
\toprule
\multicolumn{1}{l}{}                &      Encoder        &  & \multicolumn{3}{l}{Binary} &  & \multicolumn{11}{l}{Multi-Class}                                                                  \\
\cmidrule{4-6} \cmidrule{8-18}
\multicolumn{7}{l}{} & \multicolumn{5}{l}{Panoptic Quality} &  & \multicolumn{5}{l}{$F_1-score$}\\
\cmidrule{8-12} \cmidrule{14-18}
\multicolumn{1}{l}{}                &       &  & bPQ     & mPQ     & F1     &  & $PQ^N$ & $PQ^E$ & $PQ^I$ & $PQ^C$ & $PQ^D$ &  & $F^N_1$ & $F^E_1$ & $F^I_1$ & $F^C_1$ & $F^D_1$ \\
\midrule
\multirow{7}{*}{\rotatebox[origin=c]{90}{No Reparameterized}} & FastViT-T8   &  & 0.649 & 0.477 & 0.822 &  & 0.563      & 0.546      & 0.407        & 0.399      & 0.135 &  & 0.687      & 0.697      & 0.577        & 0.519      & 0.349 \\
                                    & FastViT-T12  &  & 0.653 & 0.484 & 0.823 &  & 0.569      & 0.558      & 0.417        & 0.404      & 0.145 &  & 0.696      & 0.716      & 0.575        & 0.523      & 0.350 \\
                                    & FastViT-S12  &  & 0.653 & 0.485 & 0.824 &  & 0.573      & 0.561      & 0.412        & 0.408      & 0.128 &  & 0.697      & 0.718      & 0.575        & 0.521      & 0.358 \\
                                    & FastViT-SA12 &  & 0.654 & 0.482 & 0.824 &  & 0.568      & 0.562      & 0.412        & 0.403      & 0.142 &  & 0.696      & 0.715      & 0.581        & 0.525      & 0.366 \\
                                    & FastViT-SA24 &  & 0.658 & 0.488 & 0.825 &  & 0.575      & 0.565      & 0.417        & 0.408      & 0.135 &  & 0.701      & 0.724      & 0.579        & 0.524      & 0.341 \\
                                    & FastViT-SA36 &  & 0.660 & 0.490 & 0.827 &  & 0.574      & 0.575      & 0.417        & 0.410      & 0.125 &  & 0.704      & 0.730      & 0.582        & 0.523      & 0.367 \\
                                    & FastViT-MA36 &  & 0.659 & 0.493 & 0.826 &  & 0.579      & 0.577      & 0.420        & 0.413      & 0.132 &  & 0.706      & 0.730      & 0.584        & 0.529      & 0.367 \\
\midrule
\multirow{7}{*}{\rotatebox[origin=c]{90}{Reparameterized}}    & FastViT-T8   &  & 0.649 & 0.477 & 0.822 &  & 0.563      & 0.548      & 0.408        & 0.401      & 0.141 &  & 0.701      & 0.725      & 0.581        & 0.526      & 0.353 \\
                                    & FastViT-T12  &  & 0.653 & 0.484 & 0.823 &  & 0.570      & 0.561      & 0.414        & 0.399      & 0.150 &  & 0.696      & 0.716      & 0.575        & 0.523      & 0.350 \\
                                    & FastViT-S12  &  & 0.654 & 0.485 & 0.824 &  & 0.572      & 0.562      & 0.416        & 0.406      & 0.137 &  & 0.697      & 0.718      & 0.575        & 0.521      & 0.358 \\
                                    & FastViT-SA12 &  & 0.654 & 0.482 & 0.824 &  & 0.567      & 0.554      & 0.417        & 0.398      & 0.145 &  & 0.696      & 0.715      & 0.581        & 0.525      & 0.366 \\
                                    & FastViT-SA24 &  & 0.658 & 0.488 & 0.825 &  & 0.574      & 0.566      & 0.418        & 0.406      & 0.149 &  & 0.701      & 0.724      & 0.579        & 0.524      & 0.341 \\
                                    & FastViT-SA36 &  & 0.660 & 0.490 & 0.827 &  & 0.575      & 0.569      & 0.431        & 0.407      & 0.138 &  & 0.704      & 0.730      & 0.582        & 0.523      & 0.367 \\
                                    & FastViT-MA36 &  & 0.659 & 0.493 & 0.826 &  & 0.577      & 0.571      & 0.417        & 0.413      & 0.145 &  & 0.706      & 0.730      & 0.584        & 0.529      & 0.367  \\
\bottomrule
\end{tabular}%
}
\end{table*}
We used the PanNuke dataset to evaluate the performance of our models on nuclei instance segmentation and classification of five distinct cell types: neoplastic, epithelial, inflammatory, connective, and dead cells. We consider the PQ and F-score for each nuclei type and the F-score for detection to perform a robust analysis.
Furthermore, we compare our results with DIST, MASK-RCNN, MICRO-Net, HoVer-UNet, CellViT256, and CellViT-SAM-H. In the following results, we consider three NuLite versions: NuLite-T, NuLite-M, and NuLite-H, which respectively use FastViT-S12, FastViT-SA36, and FastViT-MA36 as encoders and, in inference, they are reparameterized as described in \cite{vasu2023fastvit}. In the ablation study section, we justify the selected encoders and reparameterization.
To compare our model with others, we take into account three aspects. First, we analyze the PQ for each nucleus type, reported in \ref{tab:pq_pannuke}, then we analyze the F1-score, reported in Table \ref{tab:f1_pannuke}. Lastly, we analyze the results regarding binary Panoptic Quality (bPQ) and multi-class Panoptic Quality (mPQ) over each tissue. All results are an average of over three training sessions, as the authors' dataset suggested.

As notable, in Table \ref{tab:pq_pannuke}, the CellViT-SAM-H model outperforms all models. However, our NuLite versions follow closely, particularly excelling in Inflammatory, where NuLite-M has the highest values with a PQ of 0.4373. Overall, NuLite models, particularly NuLite-H, show competitive results for PQ metrics that outperform all other models.
In binary detection, Table \ref{tab:f1_pannuke}, the CellViT-SAM-H model and the NuLite-M and H models achieve the highest F1-scores of 0.83, showing strong detection capabilities. The NuLite-H model performs exceptionally well for Neoplastic and Epithelial nuclei, with F1-scores of 0.71 and 0.73, respectively, aching the SOTA results.
The CellViT256 and NuLite models show strong results in nucleus types like inflammatory, connective, and dead nuclei. Notably, NuLite-H achieves top F1 Scores in classifying Connective nuclei, while NuLite-M and NuLite-H excel in the Dead nuclei category. Despite the CellViT-SAM-H model being the best model performer, NuLite models, especially NuLite-H, perform almost equally. They often rank second and show strengths in specific areas like binary detection and classification of more challenging nuclei types.
Lastly, we also compared our versions with CellViT over tissue types, as shown in Table \ref{tab:tissue_res}. Again, these results demonstrate that NuLite in all its versions is similar to CellViT in binary panoptic quality (bPQ) and multi-class panoptic quality (mPQ). 
Therefore, these results show that using NuFast-ViT-H, we practically obtain the same results as CellViT-SAM-H in terms of PQ and F1-score. Also, considering a tiny version of NuLite, the results are not different from CellViT-SAM-H and are better or equal to CellViT256, which is the lightest version of it.

\begin{table*}[h!]
\centering
\caption{Comparison between CellViT and NuLite (ours) over two input shapes (256, 1024), in terms of the number of parameters, number of multiplications and additions, and estimated size GPU latency}
\label{tab:complexity}
\resizebox{\textwidth}{!}{%
\begin{tabular}{lllllllllllll}
\toprule
\multicolumn{1}{l}{Model}   & Encoder      &  & \#    Parameters (M) &  & \multicolumn{2}{l}{GLOPS}                          & \multicolumn{1}{l}{} & \multicolumn{2}{l}{Estimated Total Size   (MB)}    & \multicolumn{1}{c}{} & \multicolumn{2}{l}{GPU Latency (ms)}               \\
\cmidrule{6-7} \cmidrule{9-10} \cmidrule{12-13}
\multicolumn{1}{l}{}        &              &  &                      &  & \multicolumn{1}{l}{256} & \multicolumn{1}{l}{1024} & \multicolumn{1}{c}{} & \multicolumn{1}{l}{256} & \multicolumn{1}{c}{1024} & \multicolumn{1}{c}{} & \multicolumn{1}{l}{256} & \multicolumn{1}{l}{1024} \\
\midrule
\multirow{2}{*}{CellViT}    & ViT-256      &  & 46.75                &  & 132.89                  & 2,125.94                 &                      & 1,859.98                & 26,953.06                &                      & $35.71\pm0.37$          &   $1169.7\pm148.92$     \\
                            & SAM-H        &  & 699.74               &  & 214.20                  & 3,413.41                 &                      & 6,002.34                & 45,612.96                &                      & $103.89\pm0.97$         &   $2389.14\pm150.18$      \\
\midrule
\multirow{7}{*}{NuLite}     & FastViT-T8   &  & 5.28                 &  & 10.83                   & 173.22                   &                      & 380.01                  & 5,764.12                 &                      & $13.42\pm0.77$          & $178.89\pm18.05$         \\
                            & FastViT-T12  &  & 10.13                &  & 19.36                   & 309.70                   &                      & 528.54                  & 7,850.22                 &                      & $14.87\pm0.45$          & $214.77\pm23.66$         \\
                            & FastViT-S12  &  & 12.05                &  & 19.76                   & 316.16                   &                      & 546.18                  & 8,017.28                 &                      & $14.76\pm0.41$          & $212.3\pm21.4$           \\
                            & FastViT-SA12 &  & 14.16                &  & 19.76                   & 316.18                   &                      & 555.41                  & 8,038.31                 &                      & $14.78\pm0.83$          & $212.98\pm23.6$          \\
                            & FastViT-SA24 &  & 24.13                &  & 21.46                   & 343.22                   &                      & 715.31                  & 9,999.14                 &                      & $21.84\pm0.35$          & $267.83\pm24.81$         \\
                            & FastViT-SA36 &  & 34.10                &  & 23.15                   & 370.25                   &                      & 875.21                  & 11,959.97                &                      & $29.99\pm1.79$          & $310.44\pm24.64$         \\
                            & FastViT-MA36 &  & 47.93                &  & 32.54                   & 520.45                   &                      & 1,067.91                & 14,214.10                &                      & $33.37\pm1.34$          & $446.3\pm35.25$          \\
\midrule
\multirow{7}{*}{NuLite-Rep} & FastViT-T8   &  & 5.26                 &  & 10.82                   & 173.17                   &                      & 341.02                  & 5,141.21                 &                      & $9.11\pm0.54$           & $159.97\pm18.11$         \\
                            & FastViT-T12  &  & 10.09                &  & 19.35                   & 309.65                   &                      & 472.35                  & 6,952.55                 &                      & $10\pm0.27$             & $187.04\pm20.67$         \\
                            & FastViT-S12  &  & 12.01                &  & 19.76                   & 316.11                   &                      & 489.99                  & 7,119.61                 &                      & $9.96\pm0.23$           & $189.04\pm16.61$         \\
                            & FastViT-SA12 &  & 14.13                &  & 19.76                   & 316.13                   &                      & 501.34                  & 7,174.21                 &                      & $10.45\pm0.27$          & $197.35\pm19.55$         \\
                            & FastViT-SA24 &  & 24.08                &  & 21.45                   & 343.16                   &                      & 623.45                  & 8,531.03                 &                      & $14.69\pm0.86$          & $225.49\pm18.37$         \\
                            & FastViT-SA36 &  & 34.04                &  & 23.14                   & 370.20                   &                      & 745.57                  & 9,887.84                 &                      & $18.66\pm0.4$           & $266.82\pm19.13$         \\
                            & FastViT-MA36 &  & 47.85                &  & 32.53                   & 520.39                   &                      & 913.95                  & 11,753.44                &                      & $23.05\pm0.86$          & $402.67\pm30.99$        \\
\bottomrule
\end{tabular}%
}
\end{table*}
\subsection{Ablation study}
In this section, we describe the methodology used for selecting our models. We evaluated our NuLite using all versions of FastViT, considering both reparameterized and non-reparameterized variants during the inference phase. As shown in Table \ref{tab:ablation}, the results indicate that reparameterization does not significantly affect performance. Consequently, we opted for the reparameterized versions due to their superior computational efficiency, as demonstrated in Table \ref{tab:complexity}.
To provide a comprehensive representation, we selected three versions of NuLite, namely NuLite-T, NuLite-M, and NuLite-H, corresponding to tiny, medium, and large configurations, respectively. This selection was informed by the evaluation results on the PanNuke dataset, detailed in Table \ref{tab:ablation}. The metrics used for evaluation included binary panoptic quality (PQ), multi-class panoptic quality (mPQ), F1-score, panoptic quality, and F1-score for each nucleus type in the PanNuke dataset. The nucleus types are denoted as Neoplastic (N), Epithelial (E), Inflammatory (I), Connective/Soft Tissue (C), and Dead (D).
Upon analyzing the results for the reparameterized models, it is apparent that the performance metrics are closely aligned across most variants. Thus, model selection also considered inference time and model complexity, as detailed in Table \ref{tab:complexity}. First, we excluded the results with FastViT-T8 because despite being the lightest, the results were the worst. Then, we grouped by GLOPS the rest of the models and obtained three groups, namely FastViT-T12, FastViT-S12, FastViT-SA12 as tiny models, FastViT-SA24 and FastViT-SA36 for medium models, and FastViT-MA36 as a huge model.
Subsequently, we chose FastViT-S12 as the backbone for NuLite-T(iny), FastViT-SA24 as the backbone of NuList-M(medium), and FastViT-SA36 for NuLite-H(uge), each one for the best performance in its group. 

\subsection{Models complexity analysis}
To prove that our model, NuLite, has a lower complexity than CellViT, in Table \ref{tab:complexity}, we report an exhaustive comparison between NuLite and CellViT in terms of parameters count and GFlops, estimated size, and latency on GPU using an input shape of 256 and 1024. In particular, we consider all FastViT models (T8, T12, S12, SA24, SA36, and MA36) and the reparameterized versions in the inference step. Instead, for CellViT, we use the version with ViT256 and SAM-H as encoders. 
Results concerning GFlops and estimated size refer to a batch with just one image. Instead, GPU latency refers to a batch size of 4, and we repeated the experiments 100 times and reported the mean and variance in milliseconds. We conducted the measure on a server with AMD EPYC 7282 16-Core Processor, RAM 64 GB, and GPU Nvidia Tesla V100S 32 GB.
We consider reparameterization because even if the number of parameters and GFlops are roughly the same, the inference time is lower using it in the inference step, so we limit our consideration of these results in this analysis. 
According to these results, in terms of GFLOPS, all versions of NuLite are significantly lower than CellViT; further, all NuLite sizes are lower than CellViT. In terms of parameters, CellViT with SAM as the backbone is larger than our model, but the version with ViT-256 is smaller than NuLite with FastViT-MA36 as the backbone. CellViT takes a longer GPU latency than our NuLite versions because the amount of multiplication and addition is smaller than all CellViT versions. Moreover, if we consider our worst NuLite GPU latency for each shape, namely using FastViT-MA36 without reparameterization, it is faster than the best case of CellViT, namely CellViT256; moreover, it is almost two times faster with shape $1024\times1024$. Furthermore, another critical aspect is the estimated total size, which indicates the amount of memory used by the model during an inference step on a batch size of one, where our NuLite models outperform all CellViT models.
Limiting the analysis only to selected NuLite, we compared them with CellViT256 and CellViT-SAM-H in terms of the parameters, GFLOPS, GPU latency, and estimated size. 

\begin{table}[h]
\caption{Speedup of NuLite compared to CellViT}
     \label{tab:speed_up}
    \begin{subtable}[h]{0.49\columnwidth}
        \centering
        \caption{\#Parameters}
        \label{tab:speed_up_params}
        \resizebox{\linewidth}{!}{%
            \begin{tabular}{lll}
                \toprule
                         & CellViT-256  & CellViT-SAM-H \\
                \midrule
                NuLite-T & $3.89\times$ & $58.27\times$ \\
                NuLite-M & $1.37\times$ & $20.56\times$ \\
                NuLite-H & $0.98\times$ & $14.62\times$ \\
                \bottomrule
            \end{tabular}%
        }
    \end{subtable}
    \hfill
    \begin{subtable}[h]{0.49\columnwidth}
        \centering
        \caption{GFLOPS}
        \label{tab:speed_up_gflops}
        \resizebox{\linewidth}{!}{%
            \begin{tabular}{lll}
                \toprule
                         & CellViT-256  & CellViT-SAM-H \\
                \midrule
                NuLite-T & $6.73\times$ & $10.84\times$ \\
                NuLite-M & $5.74\times$ & $9.26\times$  \\
                NuLite-H & $4.08\times$ & $6.58\times$ \\
                \bottomrule
            \end{tabular}%
        }
    \end{subtable}
\end{table}
Table \ref{tab:speed_up} presents a comparative analysis of the number of parameters for the NuLite models against CellViT architectures, including CellViT-256 and CellViT-SAM-H. The values indicate how parameter-rich the CellViT models are to the corresponding NuLite variants. CellViT-256 has 3.89 to 0.89 times the number of parameters of NuLite, whereas CellViT-SAM-H exhibits a significantly larger increase, from 14.62 to 58.27 times more than NuLite models.
In the same way, Table \ref{tab:speed_up_gflops} presents the comparative analysis for GFLOPS; the CellViT-256 model is 4.08 to 6.73 times more intensive than NuLite models, while CellViT-SAM-H ranges from 6.58 to 10.84 times more.
\begin{table}[h!]
\centering
\caption{Estimated size speedups for NuLite models with input size $256\times256$ pixels with overlap 64 pixels and input size $1024 \times 1024$ pixel against CellViT models.}
\label{tab:size_speed_up}
\resizebox{\linewidth}{!}{%
\begin{tabular}{lllllll}
\toprule
 &  & \multicolumn{2}{c}{Input Size   $256\times256$} &  & \multicolumn{2}{c}{Input Size   $1024\times1024$} \\
\cmidrule{3-4} \cmidrule{6-7}
Model         &  & CellViT-256         & CellViT-SAM-H         &  & CellViT-256          &  CellViT-SAM-H          \\
\midrule
NuLite-T &  & $3.8\times$            & $12.25\times$          &  & $3.79\times$            & $6.41\times$            \\
NuLite-M &  & $2.49\times$           & $8.05\times$           &  & $2.73\times$            & $4.61\times$            \\
NuLite-H &  & $2.04\times$           & $6.57\times$           &  & $2.29\times$            & $3.88\times$         \\
\bottomrule
\end{tabular}%
}
\end{table}
\begin{table}[!h]
\centering
\caption{Inference speedups for NuLite models with input size $256\times256$ pixels with overlap 64 pixels and input size $1024 \times 1024$ pixel against CellViT models.}
\label{tab:inference_time_comp}
\resizebox{\linewidth}{!}{%
\begin{tabular}{lllllll}
\toprule
 &  & \multicolumn{2}{c}{Input Size   $256\times256$} &  & \multicolumn{2}{c}{Input Size   $1024\times1024$} \\
\cmidrule{3-4} \cmidrule{6-7}
Model         &  & CellViT-256         & CellViT-SAM-H         &  & CellViT-256          &  CellViT-SAM-H          \\
\midrule
NuLite-T &  & $3.58\times$           & $10.43\times$          &  & $6.19\times$            & $12.64\times$           \\
NuLite-M &  & $1.91\times$           & $5.57\times$           &  & $4.38\times$            & $8.95\times$            \\
NuLite-H &  & $1.55\times$           & $4.51\times$           &  & $2.9\times$             & $5.93\times$   \\
\bottomrule
\end{tabular}%
}
\end{table}
\begin{figure}[!h]
    \centering
    \includegraphics[width=\linewidth]{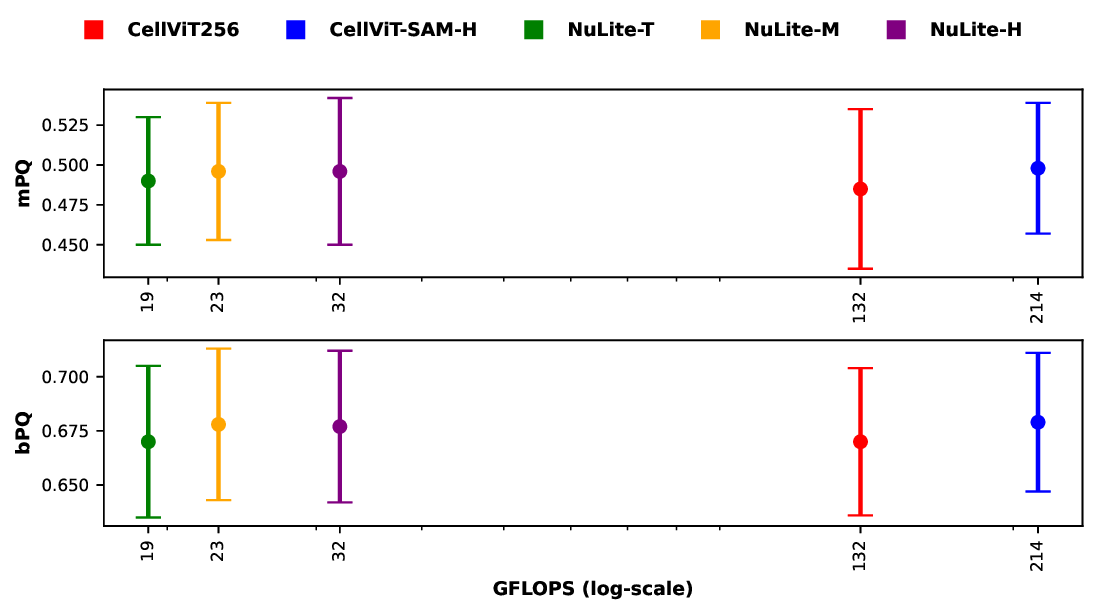}
    \caption{Comparison between CellViT256, CellViT-SAM-H, NuLite-T, NuLite-M, and NuLite-H in terms of multi-class and binary panoptic quality related to GFLOPS expressed in giga.}
    \label{fig:conf_time_pq}
\end{figure}
Again, Table \ref{tab:size_speed_up} presents the comparative analysis for estimated size during the inference step; for input size $256\times256$, the CellViT-256 model consumes from 2.04 to 3.8 times more memory than NuLite models, while CellViT-SAM-H ranges from 6.57 to 12.25 times more; for input size $1024\times1024$, the CellViT-256 model consumes from 2.29 to 3.79 times more memory than NuLite models, while CellViT-SAM-H ranges from 3.88 to 6.41 times more.

\begin{figure*}[!ht]
    \begin{minipage}[b]{\linewidth}
        \centering
        \includegraphics[width=\linewidth]{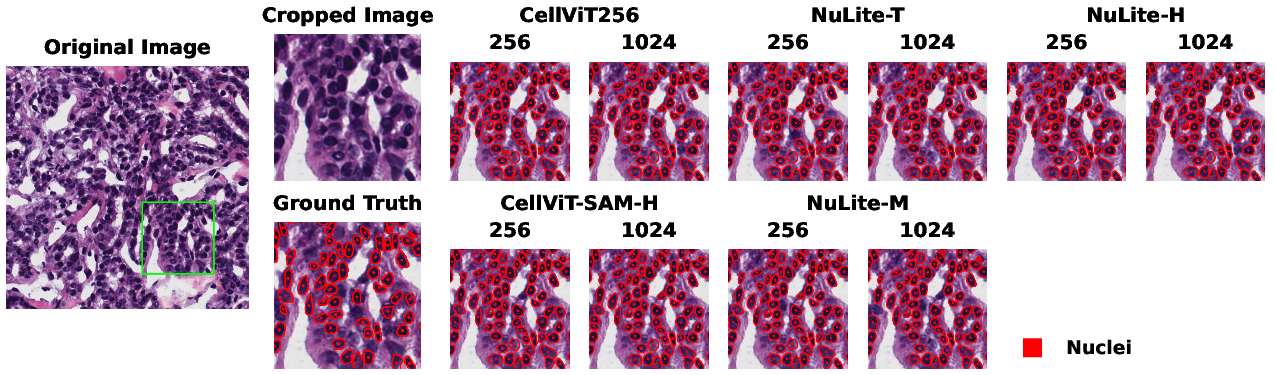}
        \caption{Segmentation masks generated by CellViT256, NuLite-T, NuLite-H, CellViT-SAM-H, and NuLite-M models on a histological image of tissue from the MoNuSeg dataset. Models were evaluated at $256\times256$ and $1024\times1024$ resolutions. Masks highlight nuclei.}
        \label{fig:monuseg_res}
    \end{minipage}
    \begin{minipage}[b]{\linewidth}
        \centering
        \captionof{table}{Comparison of CellViT, CellViT-SAM-H, and NuLite models (NuLite-T, NuLite-M, NuLite-H) across MoNuSeg, CoNSeP, and GlySAC datasets. Metrics include Detection Quality (DQ), Segmentation Quality (SQ), Panoptic Quality (PQ), and detection precision ($P_d$), recall ($R_d$), and F1-score ($F_{1,d}$) for patch sizes of $256\times256$ px and $1024\times1024$ px. The best results are highlighted in bold, with the second-best in underlined text.}
        \label{tab:ext_results_binary}
        \resizebox{\linewidth}{!}{%
        \begin{tabular}{clllllllllllllllll}
        \toprule
        \multicolumn{1}{l}{}     &         Model      &  & \multicolumn{7}{l}{Patch-Size: $256\times256$ px - Overlap: 64 px}                                  & \multicolumn{1}{c}{} & \multicolumn{7}{l}{Patch-Size: $1024\times1024$ px}                                   \\
        \cmidrule{4-10} \cmidrule{12-18}
        \multicolumn{1}{l}{}     &        &  & DQ             & SQ             & PQ             &  & $P_d$          & $R_d$          & $F_{1,d}$        &                      & DQ             & SQ             & PQ             &  & $P_d$          & $R_d$          & $F_{1,d}$        \\
        %\cmidrule{4-6} \cmidrule{8-10} \cmidrule{12-14} \cmidrule{16-18}
        \midrule
        \multirow{5}{*}{\rotatebox[origin=c]{90}{MoNuSeg}} & CellViT-256   &  & 0.861          & 0.771          & 0.664          &  & 0.830          & 0.869          & 0.848          &                      & {\ul 0.868}    & 0.771          & {\ul 0.670}    &  & 0.839          & 0.859          & 0.848          \\
                                 & CellViT-SAM-H &  & \textbf{0.869} & \textbf{0.775} & \textbf{0.674} &  & \textbf{0.850} & {\ul 0.886}    & {\ul 0.867}    &                      & \textbf{0.872} & \textbf{0.778} & \textbf{0.678} &  & \textbf{0.855} & \textbf{0.893} & \textbf{0.873} \\
                                 & NuLite-T      &  & 0.859          & 0.771          & 0.663          &  & {\ul 0.848}    & \textbf{0.900} & \textbf{0.872} &                      & 0.861          & 0.771          & 0.664          &  & {\ul 0.842}    & {\ul 0.879}    & {\ul 0.859}    \\
                                 & NuLite-M      &  & {\ul 0.865}    & {\ul 0.774}    & {\ul 0.670}    &  & 0.825          & 0.868          & 0.845          &                      & 0.863          & {\ul 0.775}    & 0.669          &  & 0.833          & 0.867          & 0.849          \\
                                 & NuLite-H      &  & 0.864          & \textbf{0.775} & {\ul 0.670}    &  & 0.841          & 0.876          & 0.858          &                      & 0.862          & {\ul 0.775}    & 0.668          &  & 0.841          & 0.858          & 0.848          \\
        \midrule
        \multirow{5}{*}{\rotatebox[origin=c]{90}{CoNSeP}}  & CellViT-256   &  & 0.668          & 0.757          & 0.507          &  & 0.779          & 0.696          & 0.731          &                      & 0.665          & 0.759          & 0.507          &  & 0.780          & 0.712          & 0.740          \\
                                 & CellViT-SAM-H &  & \textbf{0.706} & \textbf{0.776} & \textbf{0.549} &  & {\ul 0.817}    & {\ul 0.712}    & 0.757          &                      & \textbf{0.714} & 0.771          & \textbf{0.552} &  & 0.793          & \textbf{0.766} & \textbf{0.775} \\
                                 & NuLite-T      &  & 0.677          & 0.763          & 0.518          &  & 0.785          & 0.694          & 0.732          &                      & 0.681          & 0.763          & 0.521          &  & 0.758          & 0.686          & 0.716          \\
                                 & NuLite-M      &  & 0.695          & 0.770          & 0.537          &  & \textbf{0.836} & \textbf{0.717} & \textbf{0.768} &                      & {\ul 0.707}    & {\ul 0.772}    & {\ul 0.547}    &  & \textbf{0.824} & {\ul 0.731}    & {\ul 0.771}    \\
                                 & NuLite-H      &  & {\ul 0.697}    & {\ul 0.771}    & {\ul 0.539}    &  & 0.815          & \textbf{0.717} & {\ul 0.759}    &                      & 0.705          & \textbf{0.773} & {\ul 0.547}    &  & {\ul 0.807}    & {\ul 0.731}    & 0.763          \\
        \midrule
        \multirow{5}{*}{\rotatebox[origin=c]{90}{GlySAC}}  & CellViT-256   &  & \textbf{0.753} & \textbf{0.743} & \textbf{0.564} &  & 0.836          & 0.810          & {\ul 0.820}    &                      & \textbf{0.751} & \textbf{0.744} & \textbf{0.563} &  & 0.835          & {\ul 0.811}    & 0.819          \\
                                 & CellViT-SAM-H &  & {\ul 0.748}    & {\ul 0.742}    & {\ul 0.561}    &  & {\ul 0.842}    & \textbf{0.815} & \textbf{0.825} &                      & 0.745          & 0.742          & 0.558          &  & \textbf{0.852} & 0.808          & {\ul 0.827}    \\
                                 & NuLite-T      &  & {\ul 0.748}    & 0.741          & 0.560          &  & 0.826          & 0.797          & 0.809          &                      & {\ul 0.748}    & {\ul 0.743}    & {\ul 0.561}    &  & {\ul 0.849}    & 0.805          & 0.823          \\
                                 & NuLite-M      &  & 0.744          & 0.735          & 0.552          &  & \textbf{0.843} & {\ul 0.813}    & \textbf{0.825} &                      & 0.743          & 0.735          & 0.552          &  & 0.845          & \textbf{0.823} & \textbf{0.830} \\
                                 & NuLite-H      &  & 0.746          & 0.740          & 0.558          &  & 0.825          & 0.803          & 0.811          &                      & \textbf{0.751} & 0.739          & 0.560          &  & 0.846          & {\ul 0.811}    & 0.826             \\
                    \bottomrule
        \end{tabular}%
        }
        \end{minipage}
\end{figure*}

Table \ref{tab:inference_time_comp} compares the inference speedups of NuLite models relative to CellViT architectures, specifically CellViT-256 and CellViT-SAM-H, for different patch sizes ($256\times256$ and $1024\times1024$ pixels). The values indicate how much faster the NuLite models perform than the CellViT models. For a $256\times256$ patch size, the NuLite models speed up from 1.55 to 3.58 times CellViT-256 and from 4.51 to 10.43 times over CellViT-SAM-H. For a $1024\times1024$ patch size, the NuLite models speed up from 2.9 to 6.19 times CellViT-256 and from 5.93 to 12.64 times over CellViT-SAM-H. 
Lastly, to prove that our models are lighter than CellViT but performing as well as it, Figure \ref{fig:conf_time_pq} shows a comparison between them in terms of mPQ and bPQ, each error bar is the average standard deviation over tissue, on the x-axis there are GPLOS on log-scale. Analyzing this image, we can note that our models are less complex than CellViT variants, but the results are approximately the same. These aspects indicate that NuLite models maintain a lower computational and parameter footprint than highly demanding SOTA architectures, emphasizing their efficiency. Furthermore, the consistent performance advantage highlights the efficiency of NuLite models in inference speed, mainly when dealing with larger image patches. 
\begin{figure*}[!ht]
   \begin{minipage}[b]{\linewidth}
    \centering
    \includegraphics[width=\linewidth]{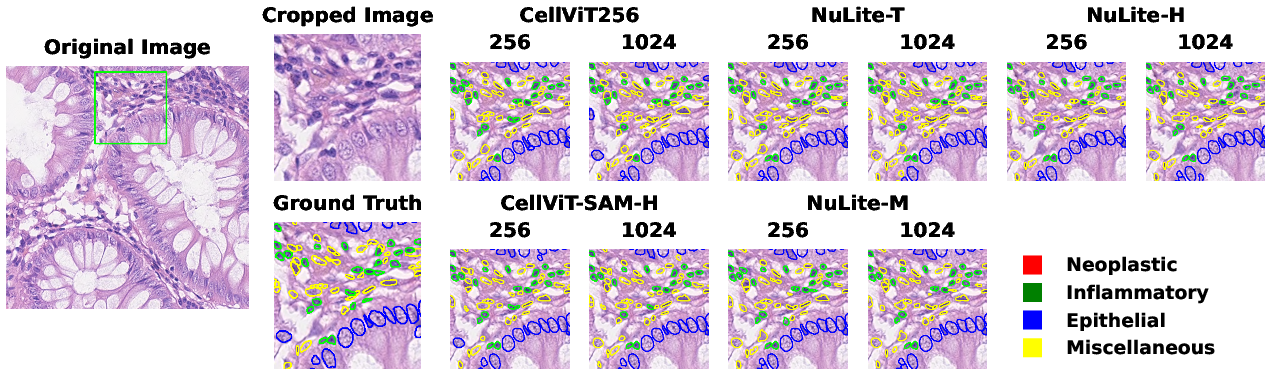}
    \caption{Segmentation masks generated by CellViT256, NuLite-T, NuLite-H, CellViT-SAM-H, and NuLite-M models on a histological image of epithelial tissue from the CoNSeP dataset. Models were evaluated at $256\times256$ and $1024\times1024$ resolutions. Masks highlight neoplastic, inflammatory, epithelial, and miscellaneous regions.}
    \label{fig:consep_res}
\end{minipage}
\begin{minipage}[b]{\linewidth}
\centering
\captionof{table}{Performance of CellViT-256, CellViT-SAM-H, NuLite-T, NuLite-M, and NuLite-H on the CoNSeP dataset across Neoplastic, Epithelial, Inflammatory, and Miscellaneous nuclei type with two patch sizes ($256\times256$ px and $1024\times1024$ px). Metrics include Detection Quality (DQ), Segmentation Quality (SQ), and Panoptic Quality (PQ), along with detection precision ($P_d$), recall ($R_d$), and F1-score ($F_{1,d}$). The best results are highlighted in bold, and the second-best results are underlined.}
\label{tab:consep_tab}
\resizebox{\linewidth}{!}{%
\begin{tabular}{lllllllllllllllllllllllllllllllll}
\toprule
Model             &  & \multicolumn{15}{l}{Patch-Size: $256\times256$ px - Overlap: 64}                                                                          &  & \multicolumn{15}{l}{Patch-Size: 1024 x   1024 px}                                                                         \\
\cmidrule{3-17} \cmidrule{19-33}
&  & \multicolumn{3}{l}{Neoplastic} &  & \multicolumn{3}{l}{Epithelial} &  & \multicolumn{3}{l}{Inflammatory} &  & \multicolumn{3}{l}{Miscellaneous} &  & \multicolumn{3}{l}{Neoplastic} &  & \multicolumn{3}{l}{Epithelial} &  & \multicolumn{3}{l}{Inflammatory} &  & \multicolumn{3}{l}{Miscellaneous} \\
\cmidrule{3-5} \cmidrule{7-9} \cmidrule{11-13} \cmidrule{15-17} \cmidrule{19-21} \cmidrule{23-25} \cmidrule{27-29} \cmidrule{31-33}
       &  & DQ       & SQ       & PQ       &  & DQ       & SQ       & PQ       &  & DQ        & SQ        & PQ       &  & DQ        & SQ        & PQ        &  & DQ       & SQ       & PQ       &  & DQ       & SQ       & PQ       &  & DQ        & SQ        & PQ       &  & DQ        & SQ        & PQ        \\
\midrule
\multicolumn{2}{l}{CellViT-256}   & 0.53            & 0.659          & 0.402          &  & 0.694          & 0.768          & 0.534          &  & {\ul 0.656}     & 0.801          & {\ul 0.531}    &  & 0.521           & 0.725           & 0.379          &  & 0.517           & 0.662           & 0.393          &  & 0.64           & 0.766          & 0.49           &  & 0.612           & 0.824          & 0.5            &  & 0.482           & 0.679           & 0.353          \\
\multicolumn{2}{l}{CellViT-SAM-H} & {\ul 0.562}     & \textbf{0.682} & {\ul 0.44}     &  & \textbf{0.772} & \textbf{0.79}  & \textbf{0.61}  &  & 0.635           & {\ul 0.825}    & 0.52           &  & \textbf{0.565}  & {\ul 0.748}     & \textbf{0.423} &  & 0.563           & 0.673           & 0.435          &  & \textbf{0.761} & \textbf{0.789} & \textbf{0.601} &  & \textbf{0.662}  & 0.825          & \textbf{0.546} &  & \textbf{0.583}  & \textbf{0.75}   & \textbf{0.438} \\
NuLite-T           &           & 0.547           & 0.666          & 0.418          &  & 0.747          & 0.772          & 0.577          &  & 0.619           & \textbf{0.828} & 0.514          &  & 0.548           & 0.735           & 0.403          &  & 0.543           & 0.666           & 0.416          &  & 0.737          & 0.759          & 0.56           &  & 0.566           & \textbf{0.832} & 0.471          &  & 0.545           & 0.735           & 0.401          \\
NuLite-M & & \textbf{0.571}  & {\ul 0.674}    & \textbf{0.442} &  & 0.757          & {\ul 0.779}    & 0.59           &  & \textbf{0.665}  & \textbf{0.828} & \textbf{0.549} &  & 0.553           & 0.747           & 0.412          &  & {\ul 0.587}     & {\ul 0.674}     & {\ul 0.454}    &  & {\ul 0.755}    & {\ul 0.784}    & {\ul 0.592}    &  & {\ul 0.653}     & {\ul 0.827}    & {\ul 0.539}    &  & {\ul 0.576}     & {\ul 0.748}     & {\ul 0.431}    \\
NuLite-H & & \textbf{0.571}  & {\ul 0.674}    & \textbf{0.442} &  & {\ul 0.766}    & 0.776          & {\ul 0.595}    &  & 0.596           & 0.761          & 0.487          &  & {\ul 0.559}     & \textbf{0.751}  & {\ul 0.419}    &  & \textbf{0.588}  & \textbf{0.676}  & \textbf{0.456} &  & {\ul 0.755}    & 0.783          & 0.591          &  & 0.58            & 0.705          & 0.476          &  & 0.554           & \textbf{0.75}   & 0.415          \\
\midrule
               &           & $P_d$           & $R_d$          & $F-1_d$        &  & $P_d$          & $R_d$          & $F-1_d$        &  & $P_d$           & $R_d$          & $F-1_d$        &  & $P_d$           & $R_d$           & $F-1_d$        &  & $P_d$           & $R_d$           & $F-1_d$        &  & $P_d$          & $R_d$          & $F-1_d$        &  & $P_d$           & $R_d$          & $F-1_d$        &  & $P_d$           & $R_d$           & $F-1_d$        \\
\midrule
CellViT-256 &   & 0.586           & 0.57           & 0.575          &  & 0.542          & 0.626          & 0.58           &  & 0.598           & \textbf{0.643} & 0.561          &  & 0.623           & 0.507           & 0.549          &  & 0.534           & 0.572           & 0.55           &  & 0.499          & {\ul 0.708}    & 0.577          &  & 0.61            & 0.517          & 0.522          &  & 0.612           & 0.455           & 0.511          \\
CellViT-SAM-H & & {\ul 0.642}     & 0.575          & 0.603          &  & 0.618          & 0.655          & 0.636          &  & 0.597           & \textbf{0.643} & 0.572          &  & 0.671           & {\ul 0.524}     & 0.585          &  & 0.56            & \textbf{0.621}  & 0.586          &  & \textbf{0.686} & \textbf{0.775} & \textbf{0.727} &  & {\ul 0.612}     & {\ul 0.584}    & {\ul 0.564}    &  & 0.658           & \textbf{0.567}  & {\ul 0.6}      \\
NuLite-T &          & 0.586           & 0.564          & 0.571          &  & 0.618          & 0.658          & 0.637          &  & 0.628           & 0.539          & 0.548          &  & 0.653           & 0.517           & 0.572          &  & 0.578           & 0.527           & 0.548          &  & 0.616          & 0.678          & 0.644          &  & 0.583           & 0.448          & 0.477          &  & 0.598           & 0.519           & 0.549          \\
NuLite-M & & \textbf{0.662}  & {\ul 0.596}    & \textbf{0.623} &  & {\ul 0.665}    & {\ul 0.694}    & {\ul 0.678}    &  & \textbf{0.706}  & {\ul 0.605}    & \textbf{0.633} &  & \textbf{0.706}  & 0.517           & {\ul 0.587}    &  & \textbf{0.664}  & {\ul 0.611}     & \textbf{0.633} &  & 0.622          & 0.682          & 0.65           &  & \textbf{0.667}  & \textbf{0.621} & \textbf{0.631} &  & \textbf{0.704}  & {\ul 0.54}      & \textbf{0.605} \\
NuLite-H & & 0.622           & \textbf{0.6}   & {\ul 0.608}    &  & \textbf{0.674} & \textbf{0.696} & \textbf{0.683} &  & {\ul 0.645}     & 0.569          & {\ul 0.593}    &  & {\ul 0.694}     & \textbf{0.532}  & \textbf{0.597} &  & {\ul 0.623}     & {\ul 0.611}     & {\ul 0.616}    &  & {\ul 0.645}    & 0.694          & {\ul 0.667}    &  & 0.546           & 0.528          & 0.523          &  & {\ul 0.672}     & 0.522           & 0.582            \\
\bottomrule
\end{tabular}%
}
\end{minipage}
\end{figure*}
\subsection{Results on others datasets}
To understand the capability of generalization of NuLite, we used MoNuSeg, CoNSeP, and GlySAC datasets and compared the results against CellViT. In particular, we used GlySAC and CoNSeP to evaluate segmentation and classification performance. Instead, we used MoNuSeg to evaluate only segmentation performance because it does not provide nuclei type. 
As described in the dataset section, CoNSeP, and GlySAC have different nuclei types of PanNuke, so we aligned them to compute multi-class metrics. 
All datasets contain tiles with shape 1000x1000, following the workflow introduced in \cite{horst2024cellvit}, we resized them to $1024\times1024$ pixels. Lastly, we compared the results using the input shape of $256\times256$ pixels with an overlap of 64 pixels and $1024\times1024$ pixels. The authors in \cite{horst2024cellvit} proved that using an input shape of $1024\times1024$ does not negatively affect the results, but they analyzed what changes for multi-class metrics; in this section, we also analyze this aspect to understand if it is possible to use $1024\times1024$ pixels tile as input when we use NuLite on whole slide images.
Here, we first analyze the binary metrics; table \ref{tab:ext_results_binary} contains the Detection Quality (DQ), Segmentation Quality (SQ), Panoptic Quality (PQ), Precision ($P_d$), Recall ($R_d$), and F1-score ($F_{1,d}$) for each dataset and inference configuration. First, we can confirm that using a tile of $1024\times1024$ as input is roughly equivalent to using a tile of $256\times256$ pixels with an overlap of 64 pixels. 

For the MoNuSeg Dataset with a smaller patch size (256x256), although CellViT-SAM-H demonstrates the best overall performance, leading in DQ, SQ, PQ, and precision, our solution, NuLite-T, achieves the highest recall and F1-score, highlighting its strong detection capabilities. NuLite-H also exhibits competitive performance, closely following CellViT-SAM-H in several key metrics. When considering the larger patch size (1024x1024), while CellViT-SAM-H continues to outperform others, NuLite, particularly NuLite-T, remains highly competitive, performing closely across multiple evaluation metrics.

On the CoNSeP Dataset, with a 256x256 patch size, CellViT-SAM-H again leads in DQ, SQ, and PQ and shows the best recall. However, NuLite-M excels in precision and F1-score, underlining its superior detection accuracy. With the larger patch size of 1024x1024, CellViT-SAM-H continues to dominate most categories, especially in DQ, PQ, recall, and F1-scores. Still, NuLite-M achieves the highest precision and maintains robust performance across other metrics.

For the GlySAC Dataset with a 256x256 patch size, CellViT-256 slightly outperforms others in DQ, SQ, and PQ. Nevertheless, our NuLite-M demonstrates strong precision and F1-score, leading in precision. At the larger patch size of 1024x1024, CellViT-256 and NuLite-T exhibit the best performance in DQ and PQ, with CellViT-256 achieving the highest scores overall, while NuLite-M outperforms others in both F1-score and precision.
Concerning binary results, we also show a visual example in Figure \ref{fig:monuseg_res}; in particular, it contains a tile of an image of the MoNuSeg date and an inference example of each analyzed model and each inference configuration setting. 

\begin{figure*}[!ht]
    \begin{minipage}[b]{\linewidth}
    \centering
    \includegraphics[width=\linewidth]{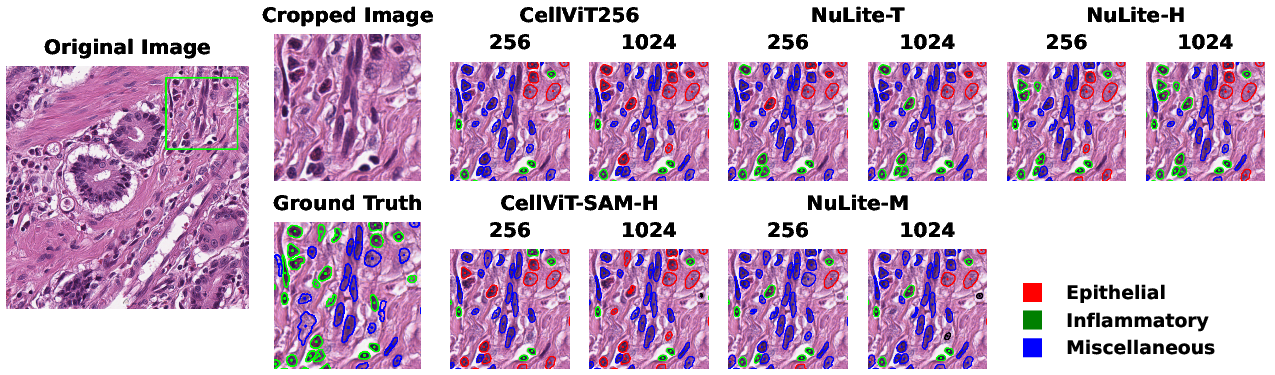}
    \caption{Comparison on different CellViT256, NuLite-T, NuLite-H, CellViT-SAM-H, and NuLite-M on an image from GlySAC dataset. Models are evaluated at different resolutions (256, 1024) and compared to ground truth. Segmentation masks highlight epithelial, inflammatory, and miscellaneous regions.}
    \label{fig:glysac_res}
\end{minipage}
\begin{minipage}[b]{\linewidth}
\centering
\captionof{table}{Performance metrics for CellViT-256, CellViT-SAM-H, NuLite-T, NuLite-M, and NuLite-H on the GlySAC dataset. Metrics are provided for different patch sizes ($256\times256$ px and $1024\times1024$ px) and include Detection Quality (DQ), Segmentation Quality (SQ), and Panoptic Quality (PQ) for Epithelial, Inflammatory, and Miscellaneous categories. Detection precision ($P_d$), recall ($R_d$), and F1-score ($F_{1,d}$) are also shown. The highest values are highlighted in bold, and the second-highest values are underlined.}
\label{tab:glysac_res}
\resizebox{\linewidth}{!}{%
\begin{tabular}{lllllllllllllllllllllllll}
\toprule
Model       &  & \multicolumn{11}{l}{Patch-Size: $256\times256$   px - Overlap: 64 }                                      &  & \multicolumn{11}{l}{Patch-Size: $1024\times1024$ px}                                     \\
\cmidrule{3-13} \cmidrule{15-25}
              &  & \multicolumn{3}{l}{Epithelial} &  & \multicolumn{3}{l}{Inflammatory} &  & \multicolumn{3}{l}{Miscellaneous} &  & \multicolumn{3}{l}{Epithelial} &  & \multicolumn{3}{l}{Inflammatory} &  & \multicolumn{3}{l}{Miscellaneous} \\
              \cmidrule{3-5} \cmidrule{7-9} \cmidrule{11-13}  \cmidrule{15-17} \cmidrule{19-21} \cmidrule{23-25}
              &  & DQ       & SQ       & PQ       &  & DQ        & SQ        & PQ       &  & DQ        & SQ        & PQ        &  & DQ       & SQ       & PQ       &  & DQ        & SQ        & PQ       &  & DQ        & SQ        & PQ        \\
\midrule
CellViT-256 &   & 0.532          & 0.722          & 0.403          &  & {\ul 0.536}     & 0.734          & {\ul 0.404}    &  & 0.306           & 0.699           & 0.217          &  & 0.51           & 0.732          & 0.388          &  & 0.5             & 0.711          & 0.378          &  & 0.285           & {\ul 0.693}     & 0.2            \\
CellViT-SAM-H &  & {\ul 0.561}    & 0.759          & {\ul 0.428}    &  & \textbf{0.549}  & \textbf{0.743} & \textbf{0.415} &  & \textbf{0.321}  & 0.696           & \textbf{0.228} &  & 0.537          & \textbf{0.767} & 0.411          &  & \textbf{0.542}  & 0.741          & \textbf{0.406} &  & \textbf{0.308}  & 0.689           & {\ul 0.217}    \\
NuLite-T          &           & 0.543          & \textbf{0.765} & 0.415          &  & 0.515           & 0.719          & 0.391          &  & {\ul 0.313}     & \textbf{0.703}  & {\ul 0.223}    &  & {\ul 0.541}    & 0.762          & {\ul 0.414}    &  & 0.5             & 0.722          & 0.38           &  & 0.298           & 0.676           & 0.214          \\
NuLite-M & & \textbf{0.562} & \textbf{0.765} & \textbf{0.431} &  & 0.525           & \textbf{0.743} & 0.398          &  & 0.306           & \textbf{0.703}  & 0.218          &  & \textbf{0.559} & {\ul 0.766}    & \textbf{0.429} &  & {\ul 0.519}     & \textbf{0.745} & {\ul 0.395}    &  & \textbf{0.308}  & \textbf{0.708}  & \textbf{0.222} \\
NuLite-H & & 0.536          & {\ul 0.764}    & 0.41           &  & 0.514           & {\ul 0.74}     & 0.391          &  & 0.297           & {\ul 0.702}     & 0.211          &  & 0.515          & 0.732          & 0.396          &  & 0.518           & {\ul 0.743}    & {\ul 0.395}    &  & {\ul 0.305}     & 0.692           & {\ul 0.217}    \\
\midrule
                      &           & $P_d$          & $R_d$          & $F-1_d$        &  & $P_d$           & $R_d$          & $F-1_d$        &  & $P_d$           & $R_d$           & $F-1_d$        &  & $P_d$          & $R_d$          & $F-1_d$        &  & $P_d$           & $R_d$          & $F-1_d$        &  & $P_d$           & $R_d$           & $F-1_d$        \\
\midrule
CellViT-256 &  & {\ul 0.519}    & 0.49           & 0.466          &  & 0.528           & \textbf{0.525} & \textbf{0.451} &  & {\ul 0.322}     & 0.331           & 0.287          &  & 0.475          & 0.488          & 0.448          &  & {\ul 0.536}     & 0.448          & 0.407          &  & 0.309           & 0.333           & 0.267          \\
CellViT-SAM-H & & 0.496          & \textbf{0.534} & {\ul 0.482}    &  & \textbf{0.545}  & {\ul 0.475}    & {\ul 0.438}    &  & \textbf{0.333}  & \textbf{0.37}   & \textbf{0.309} &  & 0.482          & \textbf{0.532} & 0.462          &  & \textbf{0.566}  & {\ul 0.452}    & {\ul 0.431}    &  & \textbf{0.349}  & {\ul 0.35}      & {\ul 0.297}    \\
NuLite-T          &           & 0.499          & 0.52           & 0.474          &  & {\ul 0.537}     & 0.413          & 0.415          &  & 0.309           & 0.349           & {\ul 0.288}    &  & {\ul 0.487}    & {\ul 0.53}     & {\ul 0.467}    &  & 0.522           & 0.386          & 0.398          &  & 0.327           & 0.331           & 0.279          \\
NuLite-M & & \textbf{0.544} & {\ul 0.523}    & \textbf{0.496} &  & 0.519           & 0.443          & 0.426          &  & 0.309           & {\ul 0.35}      & 0.286          &  & \textbf{0.561} & 0.515          & \textbf{0.501} &  & 0.505           & \textbf{0.488} & \textbf{0.433} &  & {\ul 0.341}     & \textbf{0.355}  & \textbf{0.302} \\
NuLite-H & & 0.465          & \textbf{0.534} & 0.462          &  & 0.513           & 0.452          & 0.418          &  & 0.318           & 0.309           & 0.278          &  & 0.468          & 0.525          & 0.451          &  & 0.523           & 0.441          & 0.417          &  & 0.328           & 0.326           & 0.291                 \\
\bottomrule
\end{tabular}%
}
\end{minipage}
\end{figure*}
Concerning the CoNSeP multi-class setting, we followed the alignment as shown in \cite{tommasino2023hover}; the Neoplastic class includes PanNuke's neoplastic and CoNSeP's dysplastic/malignant epithelial; the Inflammatory class encompasses PanNuke's inflammatory and CoNSeP's inflammatory; the Epithelial class consists of PanNuke's epithelial and CoNSeP's healthy epithelial; finally, the Miscellaneous class incorporates PanNuke's dead and connective tissues alongside CoNSeP's other types, which include fibroblast, muscle, and endothelial tissues.
Table \ref{tab:consep_tab} reports the results for CoNSeP multi-class comparing CellViT variants and the proposed NuLite variants across multiple tissue classes: Neoplastic, Epithelial, Inflammatory, and Miscellaneous. Each model performance is evaluated using the two configuration settings described above, with metrics such as Detection Quality (DQ), Segmentation Quality (SQ), Panoptic Quality (PQ), Precision ($P_d$), Recall ($R_d$), and F1-score ($F_{1,d}$) reported for each class.
For patch size $256\times256$ pixels with a 64-pixel overlap, the NuLite-H model shows competitive performance, tying for the top score in the Neoplastic category for both DQ (0.571) and PQ (0.442) and demonstrating strong results across other categories. CellViT-SAM-H generally leads in this setting, indicating its effectiveness with smaller image patches, particularly in the Epithelial category, where it achieves the highest DQ (0.772), SQ (0.79), and PQ (0.61) scores. However, the NuLite models, particularly NuLite-M and NuLite-H, show notable strengths in specific categories, highlighting their robustness and versatility.
When it comes to larger patch sizes, particularly $1024\times1024$ pixels, the NuLite models, especially NuLite-H, show a significant improvement and often outperform CellViT models. NuLite-H, in particular, achieves the highest PQ scores in the Neoplastic category (0.456) and ties for the highest in the Epithelial category (0.601), demonstrating its ability to maintain performance across larger contexts. NuLite-M also performs exceptionally well, achieving the highest PQ score in the Miscellaneous category (0.438) and the top DQ score in the Inflammatory category (0.665).
Regarding $F_{1,d}$ scores, NuLite-M consistently excels, particularly with larger patch sizes, achieving top scores in the Neoplastic (0.664), Epithelial (0.667), and Miscellaneous (0.605) nuclei types. This indicates that NuLite-M captures high precision while maintaining strong recall, essential for accurate and reliable classification. 
Furthermore, Figure \ref{fig:consep_res} shows an inference example for each analyzed model, highlighting a tile of an image from the CoNSeP dataset. The visual results indicate that NuLite-H and NuLite-M output the best segmentation masks, particularly in more challenging regions, underscoring their effectiveness in histological image analysis.

Concerning the results on GlySAC, we aligned the nuclei types as follows: the Epithelial class of GlySAC with Neoplatist and Epithelial of PanNuke, the Inflammatory class perfectly match, and the other class of PanNuke with the miscellaneous class of GlySAC.
The results in Table \ref{tab:glysac_res} highlight the performance differences between NuLite variants and CellViT. The key metrics evaluated are Detection Quality (DQ), Segmentation Quality (SQ), Panoptic Quality (PQ), Precision, Recall, and F1 score for each nucleus type. For \textbf{epithelial nuclei}, the NuLite-M model leads with the highest PQ score of 0.431 at the \(256\times256\) px patch size. At the \(1024\times1024\) px size, NuLite-M continues to excel with the highest PQ of 0.429, slightly outperforming both CellViT-SAM-H and NuLite-H, which also deliver strong results. Regarding \textbf{inflammatory nuclei}, CellViT-SAM-H outperforms all other models at the \(256\times256\) px size, achieving the highest PQ of 0.415. However, at the \(1024\times1024\) px patch size, NuLite-M takes the lead with a PQ of 0.395, slightly ahead of CellViT-SAM-H and NuLite-H. In the \textbf{miscellaneous nuclei} category, NuLite-M again stands out, particularly at the \(1024\times1024\) px patch size, where it achieves the highest PQ score of 0.222. This suggests that NuLite-M handles the challenging task of segmenting miscellaneous nuclei more effectively than the other models, especially with larger patches.
Examining the \textbf{F1-scores} across the models, NuLite-M consistently demonstrates strong performance, particularly in the \(1024\times1024\) px patch size, where it achieves the highest F1-scores across most categories. Notably, it reaches an F1-score of 0.501 for epithelial cells, indicating a well-balanced performance between precision and recall. CellViT-SAM-H also shows competitive performance with high F1-scores, particularly for inflammatory nuclei at the smaller patch size. Regarding precision and recall, NuLite-M has the highest precision in the epithelial and miscellaneous categories at the \(256\times256\) px patch size. In contrast, CellViT-SAM-H has the highest recall for inflammatory nuclei. This trend is consistent at the \(1024\times1024\) px patch size, where NuLite-M maintains high precision across most categories.
Lastly, Figure \ref{fig:glysac_res} shows an inference example on GlySAC, where we can observe that NuLite-H achieves good results compared to ground truth.

\section{Discussion}
\label{sec:5}
In this section, we draw back the discussion of our experimental results. According to the training results in PanNuke, we can assert that our model is equivalent to CellViT-SAM-H and, almost in every analyzed case, better than CellViT-256 in terms of each analyzed metric. Still, we can also assert that our model is less complex than CellViT, especially considering the version with SAM-H with backbone. The complexity and inference time analysis proved that our model is up to 13 times faster, with parameters up to 58 times lower, with GFLOPS up to 11 times lower, saving up to 12.25 times of memory amount during the inference.
Moreover, the comparative analysis of NuLite and CellViT across MoNuSeg, CoNSeP, and GlySAC datasets highlights several key findings related to model performance and generalization capabilities in medical image segmentation tasks. The NuLite models, especially NuLite-H, demonstrate strengths in handling larger input sizes, showing superior or competitive performance against the state-of-the-art CellViT models. These motivations make them particularly valuable for applications requiring extensive spatial analysis, such as large-scale tissue image segmentation. The consistently high scores across multiple metrics and categories underscore the versatility and robustness of NuLite-H, positioning it as a significant advancement in the field.
One notable aspect of the study was comparing performance using different patch sizes ($256\times256$ pixels with 64-pixel overlap vs. $1024\times1024$ pixels). The results demonstrate that using larger patches ($1024\times1024$ pixels) does not negatively impact the performance and may even slightly improve it in some cases. This aspect is consistent with the findings of \cite{horst2024cellvit}, which suggest that larger patch sizes can maintain or enhance the accuracy of multi-class metrics without compromising the ability of the model to delineate fine details. The slight performance variations between the two patch sizes across different models indicate that larger patches can be effectively utilized in NuLite and CellViT models, potentially simplifying the preprocessing pipeline and reducing computational overhead.
The performance metrics across MoNuSeg, CoNSeP, and GlySAC datasets indicate that CellViT and NuLite are robust in handling diverse data types. However, the NuLite models, especially the medium (NuLite-M) and high (NuLite-H) variants, consistently show competitive or superior performance in several metrics compared to CellViT. Notably, in the MoNuSeg dataset, which focuses solely on segmentation, NuLite-T achieved the highest recall (R\_d) of 0.910 with $256\times256$ patches, underscoring its ability to detect relevant instances accurately. Similarly, NuLite-H demonstrated superior recall and F1 scores in the CoNSeP dataset, which involves more complex tissue classification tasks.
The CoNSeP dataset, aligned with PanNuke nuclei types, provided a challenging environment for testing multi-class segmentation capabilities. Here, NuLite-H excelled, particularly in the Neoplastic and Miscellaneous categories, suggesting that the model is adept at handling various tissue types and complex boundaries. The strong performance in the Miscellaneous class, which includes a diverse range of tissues, further underscores the model's versatility. On the other hand, CellViT-SAM-H showed strong performance in the Epithelial class, indicating its efficacy in distinguishing epithelial tissues with high segmentation quality.
The findings from this study have important implications for using these models in whole-slide imaging (WSI) applications. The ability to effectively use larger patches ($1024\times1024$ pixels) could significantly streamline the process of analyzing large WSI data, reducing the need for extensive patch overlap and accelerating the segmentation process. 

\section{Conclusion}
\label{sec:6}
In this work, we introduced NuLite, a fast and lightweight convolutional neural network for nuclei instance segmentation and classification in H\&E stained histopathological images. With its U-Net architecture featuring one decoder and three segmentation heads for predicting nuclei, horizontal and vertical maps, and nuclei types, drawing inspiration from HoVer-Net, NuLite demonstrates considerable promise. Furthermore, we provided an extensive experimental setting on data not used for training, such as CoNSeP, MoNuSeg, and GlySAC, proving the ability to generalize our model. Therefore, our model demonstrated a state-of-the-art lightweight model in nuclei instance segmentation classification. In some scenarios, it also outperforms CellViT-SAM-H, the current SOTA, but is more complex and heavy than our NuLite.
The study reveals that NuLite, especially its medium and high variants, performs on par with or even outperforms current state-of-the-art models like CellViT. Overall, NuLite represents a significant advancement in automated medical diagnostics, offering speed and accuracy that could enhance analysis efficiency in medical contexts.
In future work, we will delve deeper into the capabilities of our model, particularly its ability to embed nuclei, as also shown in \cite{horst2024cellvit}. We aim to leverage this ability in cell-graph classification, opening up new possibilities for our model's application. Furthermore, we are committed to enhancing the entire pipeline for WSI inference.

\section*{Acknowledgments}
This study was partially supported by the PNRR MUR project PE0000013-FAIR.
We also acknowledge the CINECA award (project FVT-NSC) under the ISCRA initiative for the availability of high-performance computing resources and support.

%%Harvard
\bibliographystyle{elsarticle-num}
\bibliography{nuseg}

\end{document}